\documentclass{article}
\usepackage{multirow}
\DeclareTextFontCommand{\emph}{\em}
\usepackage{arxiv2}
\usepackage{amsthm}
\usepackage[utf8]{inputenc} % allow utf-8 input
\usepackage[T1]{fontenc}    % use 8-bit T1 fonts
\usepackage{hyperref}       % hyperlinks
\hypersetup{
  colorlinks   = true, %Colours links instead of ugly boxes
  urlcolor     = blue, %Colour for external hyperlinks
  linkcolor    = blue, %Colour of internal links
  citecolor   =  blue %Colour of citations
}
\usepackage{booktabs}       % professional-quality tables
\usepackage{amsfonts}       % blackboard math symbols
\usepackage{nicefrac}       % compact symbols for 1/2, etc.
\usepackage{microtype}      % microtypography
\usepackage{lipsum}		% Can be removed after putting your text content
\usepackage{graphicx}

\usepackage{doi}

\usepackage{amsmath}
\newtheorem{theorem}{Theorem}[section]
\newtheorem{lemma}[theorem]{Lemma}
	\newtheorem{corollary}[theorem]{Corollary}
	\newtheorem{proposition}[theorem]{Proposition}
	
	\newtheorem{definition}[theorem]{Definition}

	\theoremstyle{definition}
	
	\theoremstyle{remark}
	
	\newtheorem*{note*}{Note}
	\newtheorem{remark}[theorem]{Remark}
	\newtheorem*{remark*}{Remark}

\theoremstyle{definition}
\newtheorem{exmp}{Example}[section]
\title{The Computational Complexity of Equilibria with Strategic Constraints} %TODO Please add

\newif\ifcomments
\commentstrue
\ifcomments
\newcommand{\bruce}[1]{\textcolor{blue}{(Bruce: #1)}}
\newcommand{\koosha}[1]{\textcolor{green}{(Koosha: #1)}}
\newcommand{\changes}[1]{\textcolor{red}{(changes: #1)}}
\newcommand{\new}[1]{\textcolor{brown}{( #1)}}
\else
\newcommand{\bruce}[1]{}
\newcommand{\koosha}[1]{}
\newcommand{\changes}[1]{}
\newcommand{\new}[1]{}
\fi

\usepackage[shortlabels]{enumitem}

\author{ Bruce M. Kapron\hspace{2mm}\href{https://orcid.org/0000-0002-3295-543X}{\includegraphics[scale=0.05]{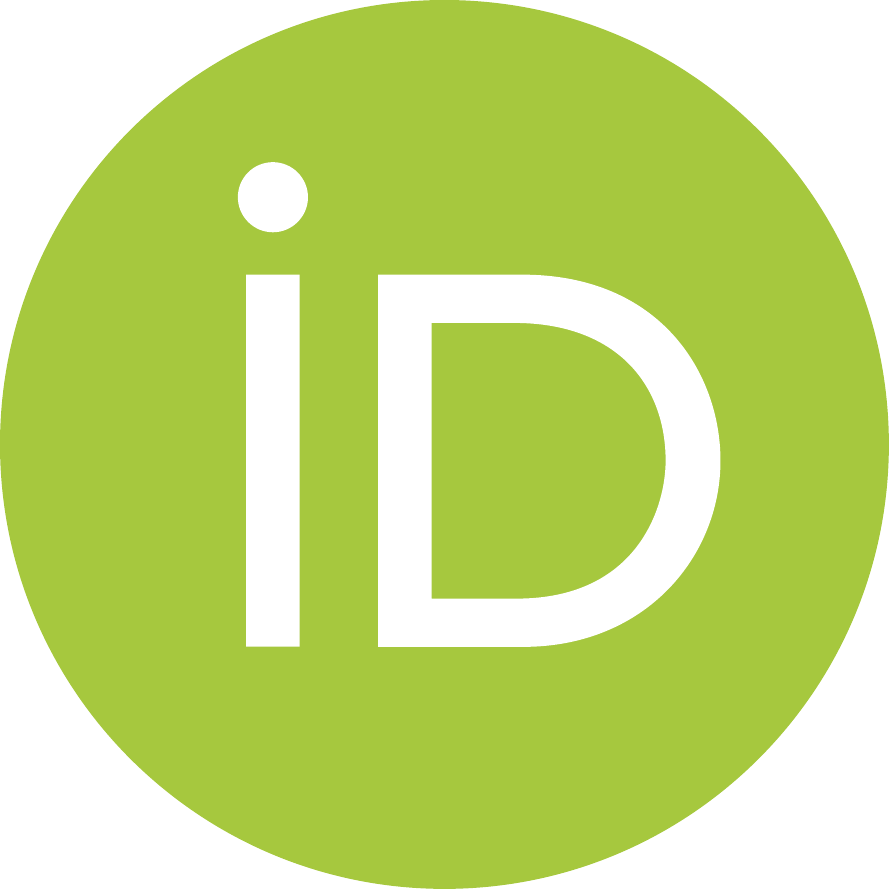}}\ \\
	Department of Computer Science\\
	University of Victoria\\
	Victoria, BC\\
	\texttt{bmkapron@uvic.ca} \\
	\And
	Koosha Samieefar\hspace{1mm}\href{https://orcid.org/0000-0001-8960-9282}{\includegraphics[scale=0.06]{orcid.pdf}} \\
	Department of Computer Science\\
	University of Victoria\\
	Victoria, BC \\
	\texttt{koosha021@gmail.com} \\
}

\begin{document}

\maketitle


\begin{abstract}

Computational aspects of solution notions such as Nash equilibrium have been extensively studied, including 
settings where the ultimate goal is to find an equilibrium that possesses some additional properties. Furthermore, in order to address issues of tractability, attention has been given to approximate versions of these problems. Our work extends this direction by considering games with constraints in which players are subject to some form of restrictions on their strategic choices.  We also consider the relationship between Nash equilibria and so-called constrained or social equilibria in this context, with particular attention to how they are related with respect to totality and complexity. Our results demonstrate that the computational complexity of finding an equilibrium varies significantly between games with slightly different strategic constraints.  In addition to examining the computational aspects of such strategic constraints, we also demonstrate that these constraints are useful for modeling problems involving strategic resource allocation and also are of interest from the perspective of behavioral game theory.
\end{abstract}

\newpage

\section{Introduction}
Game theory provides a means of mathematically modeling strategic interactions between agents, and has grown from its original application areas in the social sciences, showing itself to be relevant for modeling in computer science, particularly in areas such as networking and distributed systems. With the development of these new applications, a new focus has been put on the computational foundations of the field, going beyond the traditional mathematical existence proofs to consider issues of algorithmic feasibility. As part of this shift in focus, deep connections have been established between problems related to the existence of equilibria in games and computational complexity.

Nash equilibrium \cite{Nash0} is a central concept in game theory and much attention has been given to its computational complexity.   Nash's famous result  \cite{Nash,Nash0} proved the existence of a mixed equilibrium in any strategic game with two or more players. However, this was purely an existence result and did not address the computational complexity of the associated problem of finding a mixed equilibrium for a given game. Starting with \cite{Megido}, where it is shown that the problem is unlikely to be NP-complete, there has been a significant line of work showing investigating the hardness of this problem \cite{Chen,ComplexityNASH,Etessami,PPADcomplexity}, in particular in relation to the class TFNP of total NP search problems. This culminated in the result of \cite{Chen}, showing that the computational complexity of finding a (mixed) Nash equilibrium for two-player games is as hard as any problem in PPAD, which is a subclass of TFNP capturing the complexity of problems such as Sperner's Lemma and Brouwer's fixed-point theorem.  

In the presence of intractability, it is natural to consider relaxed notions of solutions to a problem. There are multiple constant factor approximation algorithms known for Nash equilibrium \cite{sdp,constantw,constanteDistributed,Noteconstant,13constant} however, the problem of finding an approximate solution for any additive approximation error, remains hard \cite{Chen,Inapprox,Settle}. Beyond the question of the computational complexity of finding a Nash equilibrium, we may consider equilibria in the presence of additional constraints. In this direction, there are two well-known approaches. The goal of the first approach is to find a Nash equilibrium that has certain properties (e.g., achieving a specified social welfare)  \cite{Conitzer,Gilboa,MCK1}. In the second approach, the goal is to find a constrained solution, i.e., one in which no player would deviate assuming all players only play strategies satisfying the given constraint \cite{Constrained,Social,Generalized}.

More broadly, game theory faces some foundational challenges, including how to make the concept of mixed strategies more useful in practice \cite{nopractical}. In the 1980s, the concept of mixed strategies was criticized as being "intuitively problematic", since it weakens the definition of the Nash equilibrium \cite{whymix,Ariel}.   There is no generally accepted interpretation of this concept for a variety of practical purposes. According to \cite{Corley2017}, mixed strategies are employed in games without pure Nash equilibrium, but their primary purpose is to achieve noteworthy mathematical results.  The results of this paper are motivated in part by exploring a particular interpretation of strategies as a resource, giving rise to some natural constraints that reflect this interpretation.

\subsection{Interpretation of Mixed Strategies}
  One approach to understanding mixed strategies is via a behavioral interpretation. As a central concept in economics, game theory considers scenarios where people's strategic decisions are influenced by rationality, selfishness, and utility maximization. Behavioral game theory refines this analysis by examining how people's strategic decision-making behavior is affected by their social preferences or psychological factors. The concept of mixed strategies can be interpreted in various ways \cite{Ariel}, one of which is to reinterpret a mixed equilibrium with respect to beliefs rather than actions \cite{Aumann}. A mixed strategy here models players' beliefs regarding the strategies of other players.  A second interpretation envisions players as representatives of a large number of agents. Each of the agents chooses a pure strategy, and the payoff depends on the fraction of agents choosing each strategy. Thus, the mixed strategy represents the distribution of pure strategies chosen by each population \cite{Ariel}. The paper \cite{resourcemixed}  introduces resource allocation games which provide an intuitive interpretation for a certain class of games in which equilibrium strategies represent the fraction of a resource each player allocates to each of their pure strategies. Finally, in the context of repeated games, mixed strategies may be used to represent average behavior over time. 

\subsection{Equilibrium With Strategic Constraints}

 As stated, in behavioral game theory, a player's mixed strategy may depend on psychological or other external factors. There have been a number of models that successfully explain non-equilibrium behaviors \cite{CH,GCH,Crawford,Nagel,STAHL}.  One of the main goals of this paper is to analyze a specific behavior in which all players strive to avoid overlap between their strategies and those of their opponents according to some standard measurement.

 In the extreme case, players are prohibited from using strategies used by other players, i.e., the intersection of their supports is empty. We call this problem \emph{disjoint Nash} and show that deciding whether a bi-matrix game has such an equilibrium, even under polynomial additive approximation, is NP-complete.
We also consider a stronger form of disjoint Nash, that we call \emph{partition Nash}, in which the players' supports must partition the strategy space. In Section \ref{applications} we apply this notion in the context of strategic resource allocation. Next, we will generalize disjoint Nash by allowing overlap but requiring that mixed strategies be at least $\delta$ statistically far (i.e., with respect to $L1$ distance) obtaining a  problem that we call \emph{$\delta$-far Nash}. From the perspective of seeing the mixed strategies of players as a representative of the strategies of a large number of agents,  statistical difference between the mixed strategies of the players can provide useful information for the analysis of public behavior (see Section \ref{applications}).

In general, when constraints are imposed on players' allowable strategies, it is possible to generalize Nash's notion to allow new forms of equilibria. Debreu extended the notion of equilibria to allow the possibility that a player’s set of available strategies depend on the strategic choices of other players and proved the existence of \emph{constrained (social) equilibrium} for such games when they satisfy certain assumptions \cite{Social}.  We can consider disjointness and related constraints in the setting of constrained (social) equilibrium of normal-form games where players, when evaluating a strategic response, may only consider strategies that satisfy the constraint with respect to the other players' strategies. For example, a best response for a constrained far equilibrium (see Section \ref{defs})  means that each player does not have the incentive to deviate to another mixed strategy that is statistically far enough from the opponent players. This is unlike a Nash equilibrium where a player may consider all mixed strategies, assuming all other players play a fixed strategy. In this setting, we demonstrate that disjointness has interesting computational complexity properties. To put it simply, the more we restrict the definition of Nash equilibrium with regard to farness constraints, the more difficult the problem becomes, whereas in the constrained setting, \emph{far equilibrium} will exhibit a completely different behavior.

\subsection{Main Results}

 For simplicity, we only investigate the computational aspects of two-player (bi-matrix) games. The fundamental results of the paper primarily focus on the computational complexity of two variants of Nash equilibrium, namely Nash equilibrium with statistical farness related constraints and constrained equilibrium with statistical farness related constraints. Suppose that we have a two-player game where the payoff of the players is given by a matrix that both players have $n$ possible strategies. Informally, we show that the following problems are NP-complete even under polynomial additive approximation.
\begin{itemize}
    \item $\delta$-far Nash  for any $\delta>poly(\frac{1}{n})$
    \item Disjoint Nash
    \item Partition Nash
    \item Nash equilibrium with no minor probabilities
\end{itemize}
It is natural to consider whether there is a threshold approximation error below which the problem $\delta$-far Nash becomes total (has at least one guaranteed solution). While we do not have an answer to this question, we provide a bound in which a solution is guaranteed by the existence of a Nash equilibrium. The problem of determining if a bi-matrix game has more than one equilibrium is one of the well-known Nash equilibria with constraints \cite{Gilboa,MCK1,Conitzer}. We observe that, given continuity considerations, every two-player game has an infinite number of approximate Nash equilibria. As an approximate analog, one can consider the problem of discovering two equilibrium conditions with a given statistical distance between them \cite{inapproxvariants}. We show that this problem is NP-complete even under polynomial approximation.

We also investigate the constrained version of problems that are related to farness constraints, where now the goal is to find an equilibrium such that the evaluation of a best response only considers deviations that are statistically far from those of the other player. The constrained version of disjoint Nash equilibrium may not have a solution while considering the approximate version of this problem (\emph{constrained disjoint equilibrium}) has at least one trivial solution (see \ref{polyconst}).  Consequently, we may impose an additional constraint on the strategy profile by excluding the possibility of minor probabilities (which we call \emph{restricted far equilibrium}). This problem, however, becomes NP-hard (Theorem \ref{farrestrict2}). Finally, we show that constrained far equilibrium, for some $\delta$, remains PPAD-hard and under one simple assumption, the problem will remain in PPAD (Theorems \ref{farppadhard} and \ref{farinppad}).

\subsection{Related Work}

  Problems related to Nash equilibrium with constraints have received considerable attention, and a number of results related to their complexity have been published.
  In contrast to the unconstrained setting, such problems do not necessarily admit a solution. Gilboa and Zemel's work on equilibrium with constraints such as having a guaranteed payoff or uniqueness shows that many of these problems are NP-complete \cite{Gilboa}; alternate proofs are provided in \cite{winlose,Codenotti,Conitzer,MCK1}.  In \cite{Conitzer} and \cite{MCK1} it is shown that the hardness results also hold for symmetric games. Furthermore, the hardness can be extended to symmetric and win-lose games (that is, all utilities are $0$ or $1$)  \cite{winlose}. There can potentially be a variety of constraints that can be considered and a wide range of these properties will lead to  NP-hard problems. To our knowledge, our work is the first to consider the complexity of equilibria with conditions based on a mutual constraint on players' strategies. Beyond questions of exact solutions in this setting, a limited amount of progress has been made in approximating Nash equilibrium with certain constraints unless they have a direct relationship to the social welfare of the players \cite{inapproxvariants,inapproxresults,vadhan,hardsocialwelfare}. In fact, finding a \text{PTAS} for the (unconstrained) Nash equilibrium problem has been challenging with many attempts to address the question using different approaches \cite{Query,oblivious,Communication,Quasi,spectral}. Finally, in a seminal work \cite{Settle}, it is shown that finding an approximate Nash equilibrium can have no better time complexity than a quasi-polynomial time algorithm introduced in \cite{Quasi} (assuming \text{ETH}.)

Surprisingly, considering the application of constrained games defined by Debreu \cite{Social} and Rosen \cite{rosen}  in Economics and other areas, the complexity of these problems has not received much attention until recent work \cite{concavegames}, while to our knowledge, the complexity of constrained bi-matrix games has not been considered.  Many researchers from different areas have worked on this notion which in turn explains why it has developed with a number of names in the literature including pseudo-games, social equilibrium, concave games, and common (coupled) constraint equilibrium \cite{Constrained,Generalized, Generalized2,constrainedeconomics,constrainedindef}. There are standard critiques on the notion of constrained equilibrium \cite{critic}, in particular the question of how constraints are enforced. This question may be adequately addressed in a variety of settings, e.g., correlated equilibrium or games with a mediator, and so the notion remains fundamentally interesting.

 Furthermore, works on finding constrained equilibria focus on the case of convex constraints, where solutions are guaranteed to exist by well-known mathematical facts \cite{100challenge,Social,concavegames,rosen}. It is easy to observe that our problem obviously does not fall into this setting, however, a solution is guaranteed for one of the main problems. There are a variety of works done on constrained (i.e., social or generalized Nash) equilibrium where the focus is on only pure strategies \cite{Constrained,Generalizedqvari,Generalized,Generalized2,concavegames}. A notable exception is constrained games introduced in \cite{Polowradzik} in which each player plays a mixture of two pure strategies from the set of admissible strategies.

\section{Preliminaries}\label{defs}

\subsection{Game Theory Definitions}

\begin{definition}
	A \emph{normal  form  game} is specified by a finite set of $k$ \emph{players}, a finite set $S=\{1,\dots,|S|=n\}$ of \emph{pure strategies} (or \emph{actions})\footnote{A more general formulation provides a strategy set $S_i$ for each player $i$. There is no essential difference between such a formulation and the one used in this paper when we focus on the notion of EXACT (i.e., not approximate) Nash equilibrium.}, and a vector $(U_1,\dots,U_k)$ of \emph{utility functions} where $U_i:S^k\rightarrow\mathbb{R}$.
 	
Any such game has a representation where $U_i$ is given by a $n^k$-dimensional \emph{payoff matrix} $A_i$.
For the case of two players, this gives rise to the standard notion of a \emph{bi-matrix game.}

\end{definition}	

In many situations, it is useful to allow players to randomize over pure strategies. This leads to the following:

\begin{definition}

	 Let $\triangle_{n}$ be the $n$-dimensional probability simplex, 
	    	\[
	\triangle_{n}=\left\{\textbf{x} \in \mathbb{R}^{n} \mid x_{i} \geq 0, \forall i, \sum_{i=1}^{n} x_{i}=1\right\}.
    \]
    A \emph{mixed strategy} is an element of $\triangle_{n}$, corresponding to a distribution on strategies. If  $x$ is a mixed strategy, $\mathrm{Supp}(x)$ (\emph{support} of $x$) denotes the set elements of $S$ to which $x$ assigns positive probability.
    \end{definition}
    Note that we may view a pure strategy $s$  as a mixed strategy assigning probability $1$ to $s$.
    
	\begin{definition}
	    A \emph{pure strategy profile} is a  $k$-dimensional vector $\textbf{s}=(s_1,\dots,s_k)$ of strategies, where $s_i$ is the pure strategy played by player $i$. We define a \emph{mixed strategy profile} $\textbf{x}=(x_1,\dots,x_k)$ similarly. 
	\end{definition}
If $\textbf{y}=(y_1,\dots,y_k)$ is a strategy profile and $x$ is a strategy, we use the standard notation $(x,\textbf{y}_{-i})$ to denote $(y_1,\dots,y_{i-1},x,y_{i+1},\dots,y_k)$.

\begin{definition}
	\label{expected}
	Given a pure strategy profile $\textbf{s}$, the \emph{payoff} for player $i$ is $U_i(\textbf{s})$.  We can extend this notion to mixed strategies by defining $U_i(\textbf{x})$ for  a mixed strategy profile $\textbf{x}$ to be $\mathbb{E}_{\textbf{s}\leftarrow \textbf{x}}[U_i(\textbf{s}))]$, i.e., the expected value of $U_i(\textbf{s})$ when $\textbf{s}$ is distributed according to $\textbf{x}$.

\end{definition}

	\begin{remark}
	    	In a two-player game, for a mixed strategy profile $\textbf{z}=(x,y)$ we have:
	\[U_1(x,y)=\sum_{1 \leq i\leq n } \sum_{ 1 \leq  j \leq n} \alpha_{ij} U_1(i,j) \text{ and } U_2(x,y)=\sum_{1 \leq i\leq n } \sum_{ 1 \leq  j \leq n} \alpha_{ij} U_2(i,j),\]
 where $\alpha_{ij}=x_iy_j$.

	\end{remark}

\begin{definition}
	A pure strategy, \(s\), is a \emph{best response} for player \(i\) to a strategy profile $\textbf{x}$ if for all \(s^{\prime} \in S\) we have $	U_{i}\left(s, \textbf{x}_{-i}\right) \geq U_{i}\left(s^{\prime}, \textbf{x}_{-i}\right)$.

	We can also define a \emph{mixed best response} strategy to a mixed strategy profile in the same manner.
 \end{definition}

\begin{remark}\label{bestrep}
We note that there are equivalent versions of the definitions of best response and Nash equilibrium in which the restriction to pure strategies is relaxed, i.e., players compare the payoff using a particular mixed strategy against all alternative mixed strategies.
\end{remark}

\begin{definition}
	%Consider a game $\mathcal{G}$  with \(k\) players. 
 A strategy profile \(\textbf{x}^{*}=\left(x_{1}^{*}, x_{2}^{*}, \dots, x_{k}^{*}\right)\) is a \emph{Nash equilibrium}  
 %of the game 
 if for every player \(i\) and for all \(s \in S\), 
 %for each player \(i\):
	$
	U_{i}\left(\textbf{x}^{*} \right) \geq U_{i}\left(s, \textbf{x}_{-i}^{*}\right)
$.
	\end{definition}

\begin{definition}
    A \emph{strictly dominated strategy} is a strategy that always delivers a lower payoff compared to any alternative strategy, regardless of what strategy the opponent chooses. 
\end{definition}

\begin{definition}
	%Consider a game with \(k\) players.
 A strategy profile \(\textbf{x}^{*}=\left(x_{1}^{*}, x_{2}^{*}, \dots, x_{k}^{*}\right)\) is an \emph{$\epsilon$-approximate Nash equilibrium}
 if for every player \(i\) and	for all \(s \in S\),
	$
	U_{i}\left(\textbf{x}^{*} \right) +\epsilon\geq U_{i}\left(s, \textbf{x}_{-i}^{*}\right) 
	$. We  define \emph{$\epsilon$-best response} similarly.
	
\end{definition}

\begin{remark}
    A strictly dominated strategy cannot participate in a Nash equilibrium.  However, in an approximate Nash equilibrium, a strictly dominated strategy  can be played with a small probability based on each player's utility matrix and the approximation error.
\end{remark}

 \begin{definition} \label{socialwelfaredef}

	The \emph{social welfare} of a strategy profile $\textbf{x}$ is the sum $\sum_{i=1}^{k}U_i(\textbf{x})$ of all players' payoffs.
\end{definition}

\begin{definition}
	A \emph{fully mixed strategy} is a mixed strategy in which all pure strategies in $S$ are assigned with a positive (greater than zero) probability. 
\end{definition}

This concept has been extensively studied which explains why there are various definitions \cite{totallymixed,Complete0,Complete295,Computationcomplete,Complete2020,Completely1}.

Throughout the paper, we will refer to well-known restrictions, including  \emph{zero-sum} and \emph{symmetric} games:

\begin{definition}
	A 2-player game is a \emph{zero-sum game} if $U_{1}(s_1,s_2)=-U_{2}(s_1,s_2)$ for all $s_1,s_2 \in S $. This means that one player’s payoff is equal to the other player’s loss on any given play of the game.
\end{definition}

\begin{remark}
For a bi-matrix games $\mathcal{G}$, we can represent $U_1$ and $U_2$ by matrices $R=(r_{ij})$ and $C=(c_{ij})$ respectively where $U_1(i,j)=r_{ij}$ and  $U_2(i,j)=c_{ij}$. The matrix representation of such a game can be denoted $\mathcal{G}(R, C)$.
\end{remark}

\begin{definition}
	The game $\mathcal{G}(R, C)$ is \emph{symmetric} if the  both players have the same strategies and $R=C^{T} $. In addition, we call $x^*$  a   \emph{symmetric equilibrium} if  $(x^*,x^*)$ is a Nash equilibrium. 
\end{definition}

\subsubsection{Constrained Equilibrium Definitions}
Once we consider games with constraints on strategy profiles, there are two natural notions of equilibrium. We can still consider whether there is a Nash equilibrium that satisfies the given constraint. However, we could also modify the above definitions in a way that constrains players to only consider alternate strategies that satisfy the constraint with respect to other players' strategies. Clearly, for any game with strategic constraints, a solution of the first sort implies a solution of the second sort, but in general, the converse does not hold. The approach we follow will largely be based on \cite{Constrained} and \cite{Generalized}. We use the following definition to define constrained disjoint equilibrium in  Section \ref{problems}. In section \ref{dmg}, we will investigate the relationship thoroughly.

\begin{definition} \label{CCCdef}  Let $\mathcal{G}$ be a game and $\mathcal{R}_i$  a  relation on $\prod_{i=1}^{k-1} \Delta_{n}$, for $i\in [k]$. Each $\mathcal{R}_i$ may be viewed as a set-valued mapping, so that $\mathcal{R}_i(\textbf{x}_{-i})\subseteq \Delta_n$. Let $\mathcal{R}=\{\mathcal{R}_i~|~i\in[n]\}$. A mixed strategy profile $x^*$ is a \emph{constrained equilibrium} (generalized Nash equilibrium) for  $(\mathcal{G},\mathcal{R})$ if:
    \begin{equation}
    \label{eq:bestcommonresp}
        \forall x_i \in \mathcal{R}_i(\textbf{x}^*_{-i}), \quad\quad   U_{i}\left(\textbf{x}^{*} \right) \geq U_{i}\left(x_{i}, \textbf{x}_{-i}^{*}\right)
\end{equation}
\end{definition}
In the sequel, we will only consider the case where all the $\mathcal{R}_i$'s are the same, i.e, there is one \emph{common constraint} $\mathcal{R}$.

\begin{remark}
The definition of constrained games in \cite{concavegames,rosen} does not involve  games with mixed strategies.  Our definition may be viewed as a special case of these definitions.
\end{remark}

\subsection{Computational Complexity Definitions}

\begin{definition}
	TFNP (Total Functional Nondeterministic Polynomial Time) is the class of total search problems with solutions that are poly-time verifiable. Formally, given a poly-balanced poly-time relation $R(x,y)$ the associated  \emph{NP search problem} is the (partial) multi-valued function $Q(x)=\{y~|~R(x,y)\}$. The problem is \emph{total} if $Q(x)$ is nonempty for all $x$. The class \emph{FNP} consists of all NP search problems, while TFNP consists of all total NP search problems. A search problem $Q$ is in \emph{FP} if there is a  poly-time function $Q^\prime$ such that for all $x$, $Q^\prime(x)\in Q(x)$. The \emph{decision problem} associated with $R$ is to determine, given $x$, whether there is some $y$ such that $R(x,y)$. NP is the class of all such decision problems.
	
\end{definition}

\begin{definition}

The notion of many-one reduction is extended to total search problems  as follows:
For two total search problems $R$ and $S$ we say $R \leq_{m} S$ if there exist poly-time computable functions  $f, g$ such that for all $x,y$  if $(f(x), y) \in S$  then $(x, g(y)) \in R$.

\end{definition}

\begin{remark}
    Note that this form of reduction between search problems is equivalent to a Cook reduction with one call to the oracle.
\end{remark}

The subclass PPAD of TFNP plays a central role in the complexity of the Nash equilibrium problem. The classes are  defined as follows:
\begin{definition}
We define the \emph{end-of-the-line (EOTL)} problem as follows:
Given a directed graph $\mathcal{G}$, represented by two poly-sized circuits which return the predecessor and the successor of a node (represented in binary), where each vertex has at most one predecessor and successor and a vertex $u$ in $\mathcal{G}$ with no predecessor, find another vertex $v\neq u$ with no predecessor or no successor. A search problem is in \emph{PPAD} if it is poly-time reducible to EOTL.
	
\end{definition}

In a bi-matrix game, the goal is to find a strategy profile (pair of mixed strategies) such that the Nash conditions are satisfied. If we already know the support of these strategies, determining the relative weight of the problem can be solved in polynomial time using linear programming. This immediately gives the following:
\begin{proposition} \label{encode3}
	Any $2$-player game with rational payoffs has a rational Nash
	equilibrium where the probabilities are of a bit-length polynomial with respect to the number of strategies and bit-lengths of the payoffs.
\end{proposition}

In conclusion, when we have two players, the challenge lies in finding these two subsets. This approach does not extend beyond two-player games, and in fact, in Nash's original paper, it is shown that there are $3$-player games that can have only irrational solutions. Discussion of EXACT complexity in such cases requires the use of classes such as FIXP \cite{Etessami} and (for Nash equilibrium with additional constraints) $\exists \mathbb{R}$ \cite{BerthelsenH22}.

When considering the computational complexity of additive approximation, we need to take into account the representation of the approximation term $\epsilon$. In particular, we consider the following forms of approximation \cite{vadhan}:

\begin{definition}
   
   \phantom{xxx}
   
   \begin{itemize} 
	    \item  By \emph{EXACT} we indicate an exact solution to a  problem (i.e., $\epsilon=0$.) 
     %We usually deal with the case that we have a rational solution admitting a finite representation. 
	    \item For exponential additive approximation  or  \emph{EXP-approx}, the input $\epsilon$  is encoded as a (dyadic) rational number.
	    \item For polynomial additive approximation or \emph{POLY-approx}  the input $\epsilon=\frac{1}{k}$ is represented (in unary) as $1^k$.
	\end{itemize}
\end{definition}

\section{Equilibria with Strategic Constraints }
This section outlines the formal definitions and main theorems of problems related to equilibria with strategic constraints. We also assume that the utility matrices' entries are between $[0,1]$ unless stated. We consider the standard notion of \emph{statistical distance} (\emph{total variation distance}) between mixed strategies, viewed as probability distributions on pure strategies.

\subsection{Problem Definitions}\label{problems}

\begin{definition} \label{farness}
	We call two (mixed) strategies $x \in	\triangle_{n}$ and $y \in \triangle_{n} $   \emph{$2\delta$}-far if we have $||x -y ||_1\geq 2\delta$, where $||x -y||_1=\sum_{i=1}^{n} |x_i-y_i|$.
\end{definition}

\begin{remark}
 Note that for bi-matrix games, for the mixed strategies $x$ and $y$ for the row and the column player, we simply have $0\le||x-y||_1\le 2$.
\end{remark}

\begin{definition} \label{metric}
	A Nash equilibrium $\textbf{x}^*=(x^*_1,\dots,x^*_k)$ is called a \emph{$\delta$-far Nash equilibrium} if for all $i \ne j$, $x^*_i$ and $x^*_j$ are $2\delta$-far. The approximate version of this type of equilibrium will be denoted by $(\delta,\epsilon)$-far Nash equilibrium.
\end{definition}

A specific case of this problem can also be introduced by prohibiting all players from using the strategies employed by all other players as follows:

\begin{definition}
	{(Disjoint Nash):} 
	A strategy profile $\textbf{x}=(x_1,\dots,x_k)$ is a \emph{disjoint Nash equilibrium} if it is a Nash equilibrium and \(\bigcap_{i=1}^{k} P_{i}=\emptyset\)  where $P_i=\mathrm{Supp}(x_i)$.
	
		We say that a mixed strategy profile $(x_1,\dots,x_k)$ has a \emph{disjoint support profile} if for any player $i$ and player $j$ ($i \neq j $) have disjoint supports.

\end{definition}

\begin{remark}
    When $\delta=0$, $\delta$-far Nash is just the problem of finding a Nash equilibrium, and when $\delta=1$ it is exactly disjoint Nash equilibrium. The existence of a disjoint Nash is not necessarily guaranteed, by the example given in Table \ref{tab:example}.
\end{remark}

\begin{table}[] 
\centering
\caption{A game with no disjoint Nash equilibrium}

\label{tab:example}
{\small
\resizebox{150px}{!}{%
\begin{tabular}{lll}
               & \multicolumn{2}{l}{Player 2}                        \\ \cline{2-3} 
\multicolumn{1}{l|}{\multirow{2}{*}{Player 1}} & \multicolumn{1}{l|}{1,1} & \multicolumn{1}{l|}{0,0} \\ \cline{2-3} 
\multicolumn{1}{l|}{}                          & \multicolumn{1}{l|}{0,0} & \multicolumn{1}{l|}{0,0} \\ \cline{2-3} 
\end{tabular}%
}}
\end{table}

For the purpose of investigating one of the straightforward applications of equilibria with limited overlapping support, we define the following problem that can be derived by further constraining disjoint Nash:

\begin{definition}
	{(Partition Nash):}	A strategy profile $\textbf{x}=(x_1,\dots,x_k)$ is a \emph{partition Nash equilibrium} if it is a disjoint Nash equilibrium and 
\(\bigcup_{i=1}^{n} P_{i}= S\), where $P_i=\mathrm{Supp}(x_i)$.
	\end{definition}

We also have another variant of Nash equilibrium for which the EXACT problem was also investigated in \cite{MCK1}. This variant requires a Nash equilibrium such that all strategies cannot be played with minor probabilities.

 \begin{definition}
	{(Major Nash):}	A strategy profile is a \emph{$\theta$-major Nash equilibrium} if all strategies in the support of all players are played with probability greater than $\theta$. For the approximation version, we consider \emph{$(\theta,\epsilon)$-major Nash equilibrium}.
	\end{definition}

Next, we investigate the constrained version of the previous Nash equilibrium problems with limited overlapping support. In the constrained setting, the goal is to find an equilibrium where only deviations to strategies that respect the constraint (which restricts the usage of strategies) are considered.  For example, in the constrained version of a disjoint Nash equilibrium (which we call constrained disjoint equilibrium), a best response for a player means that this player cannot deviate from an equilibrium strategy to another mixed strategy whose support has a nonempty intersection with those of the other players. This is unlike a Nash equilibrium where a player may consider all mixed strategies, assuming all other players play a fixed strategy.

\begin{definition}
    For bi-matrix games, we can define the common constraints for the constrained versions of $\delta$-far and disjoint Nash:
 \[\mathcal{R}^{n}_\mathrm{\delta}=\{(x,y)\in \Delta_n\times\Delta_n~|~ ||x-y||_1\geq 2\delta \}\]
 \[
		 \mathcal{R}^{n}_\mathrm{disjoint}=\{(x,y)\in \Delta_n\times\Delta_n~|~ \langle x,y\rangle =0 \}\]	
where $\langle x,y\rangle$ denotes the inner product. We call an equilibrium (see Definition \ref{CCCdef}) that satisfies $\mathcal{R}^{n}_\mathrm{\delta}$, a \emph{constrained $\delta$-far equilibrium},  and an equilibrium that satisfies $\mathcal{R}^{n}_\mathrm{disjoint}$  a \emph{constrained disjoint equilibrium}.
\end{definition}

\begin{remark}
    
Assume that the column player plays $y$. For the row player (similarly for the column player), we also can the relations:
\[\mathcal{R}^{n}_\mathrm{\delta}(y)=\{x\in \Delta_n~|~ ||x-y||_1\geq 2\delta \}\]
 \[
		 \mathcal{R}^{n}_\mathrm{disjoint}(y)=\{x\in \Delta_n~|~ \langle x,y\rangle =0 \}\]	
\end{remark}
\begin{remark}
     It is easy to observe that all games having exactly two pure strategies have a constrained disjoint equilibrium where both players play one pure strategy.  The constrained strategy spaces that we have defined are not convex. As shown by Rosen \cite{rosen} a solution is guaranteed for convex constraints. Recently, it has been proved that by using a computational version of Kakutani's fixed point theorem, finding an equilibrium for a game with convex constraints is PPAD-complete \cite{concavegames}. Our results show that a solution may be guaranteed when constraints are not convex. 
\end{remark}

In the constrained setting, at least one of the problems we propose, namely approximate constrained disjoint equilibrium, is tractable (Proposition \ref{polyconst}). However, in  an approximate constrained disjoint equilibrium, one selfish player can degrade the social welfare of all other players. With this drawback in mind, we direct our attention to another type of constrained equilibrium which is defined as follows:

\begin{definition}
 A \emph{$(\theta,\delta)$-restricted far equilibrium problem} is defined as follows:
    Given a bi-matrix game $\mathcal{G}$ find a constrained $\delta$-far equilibrium such that all pure strategies played with a positive probability are played with probability greater than $\theta$.   We can relax this definition by using an additive approximation error $\epsilon$  and call it \emph{$(\theta,\delta,\epsilon)$-restricted far equilibrium}. For $\delta=1$, we call the problem \emph{$(\theta,\epsilon)$-restricted disjoint equilibrium}.
\end{definition}

\subsection{Applications}\label{applications}

To provide some motivation, we begin with two simple examples which are based on the well-known ``Bach or Stravinsky'' game.  In both games, there are two players with two strategies. In our first scenario, we have two players who must choose between two activities: a row player (Ray) and a column player (Clara). Ray prefers the first activity while Clara prefers the second one. In the first game, if they both choose the same activity, one will receive a payoff of $\frac{1}{2}$ and the other will get $1$, based on their preference and if they choose different activities, they end up in a fight and both will receive $0$. In the second game, the players get a payoff of $1$ or $\frac{1}{2}$ (based on their preference) if they employ different strategies.

\begin{table}[] 
\centering
\caption{Bach or Stravinsky and the opposite game}

\label{tab:example4}
\resizebox{290px}{!}{%
\begin{tabular}{lll}
               & \multicolumn{2}{l}{\quad Clara}                        \\ \cline{2-3} 
\multicolumn{1}{l|}{\multirow{2}{*}{Ray}} & \multicolumn{1}{l|}{1,$\frac{1}{2}$} & \multicolumn{1}{l|}{0,0} \\ \cline{2-3} 
\multicolumn{1}{l|}{}                          & \multicolumn{1}{l|}{0,0} & \multicolumn{1}{l|}{$\frac{1}{2}$,1} \\ \cline{2-3} 
\end{tabular}%
\quad
\begin{tabular}{lll}
               & \multicolumn{2}{l}{\quad  Clara}                        \\ \cline{2-3} 
\multicolumn{1}{l|}{\multirow{2}{*}{Ray }} & \multicolumn{1}{l|}{0,0} & \multicolumn{1}{l|}{1,$\frac{1}{2}$} \\ \cline{2-3} 
\multicolumn{1}{l|}{}                          & \multicolumn{1}{l|}{$\frac{1}{2}$,1} & \multicolumn{1}{l|}{0,0} \\ \cline{2-3} 
\end{tabular}%

}
\end{table}

In the first game, there are two pure Nash equilibria and both players play the same strategies. This game has a mixed Nash equilibrium $(\frac{2}{3},\frac{1}{3})$ (for the row player) and $(\frac{1}{3},\frac{2}{3})$ (for the column player) where the statistical $L1$ distance of the strategies is $\frac{2}{3}$. In addition, in the first game, we do not have a Nash equilibrium where two players choose different activities, indicating that there cannot reach to desirable payoff if both players exhibit
%irrational behavior which is being uncooperative.
uncooperative behavior. 
In the second game, we have two pure Nash equilibria in which both players play different strategies while the unique mixed strategy equilibrium is similar to the previous game. In conclusion, in this game,  both players are inclined toward using different activities (strategies) based on their preferences.

\subsubsection{Tax Cheats and Policy Design}
We first consider an example of how a far Nash equilibrium can provide useful information for policy designers. This is illustrated in two games modeling a simple tax auditing problem, given in Table \ref{tab:example3}. %can be good examples where
Here the auditor and the taxpayer have two possible strategies. The auditor's strategies are either auditing the taxpayer (strategy $1$) or not auditing the taxpayer (strategy $2$). The taxpayer can either pay the taxes truthfully (strategy $1$) or cheat (strategy $2$). These bi-matrix games have no pure Nash equilibrium. We will use two different interpretations that we mentioned previously. The first game's Nash equilibrium happens when the taxpayer pays their taxes truthfully with probability $\frac{2}{3}$ and cheats with  probability  $\frac{1}{3}$. The auditor in this equilibrium will audit the taxpayer with probability  $\frac{2}{7}$ and will not audit with  probability  $\frac{4}{7}$. We can interpret this mixed strategy by assuming that the auditor is literally randomizing over the two strategies while the numbers that are related to the taxpayer indicate the proportion of taxpayers who are paying taxes and cheating respectively. The goal of policy design could be designing a policy in which the taxpayer tries to deter cheating. In the second game with this new penalty of getting caught in place, the auditor will audit with  probability  $\frac{1}{6}$ while the tax compliance rate will not change. In other words, in the Nash equilibrium of this game, the auditor will play $(\frac{1}{6},\frac{5}{6})$, and the taxpayer will play $(\frac{2}{3},\frac{1}{3})$. This shows that if we want to change tax compliance, we need to change the auditor's payoff although the fact that we have managed to lower the audit rate which is a good thing for society. To get a better compliance rate, we can either increase the payoff of getting a cheater or lower the cost of auditing. After discussing some aspects of income tax games, we can conclude simply if a far Nash equilibrium exists, the policy is effective since we want to lower the probability of playing the same strategy.\footnote{The policy designer should follow some basic assumptions, e.g., the penalty should not be extremely high. Auditing with a high probability while the taxpayers are honest is not ideal.  For games that have multiple mixed Nash equilibria, we could possibly measure the mean of the statistical distance of the mixed strategies' of the players'.} 

\begin{table}[] 
\centering
\caption{The income tax games}
{\small
\label{tab:example3}
\resizebox{330px}{!}{%
\begin{tabular}{lll}
               & \multicolumn{2}{l}{Tax payer}                          \\ \cline{2-3} 
\multicolumn{1}{l|}{\multirow{2}{*}{Auditor}} & \multicolumn{1}{l|}{2,0} & \multicolumn{1}{l|}{4,-10} \\ \cline{2-3} 
\multicolumn{1}{l|}{}                          & \multicolumn{1}{l|}{4,0} & \multicolumn{1}{l|}{0,4} \\ \cline{2-3} 
\end{tabular}%
\quad
\begin{tabular}{lll}
               & \multicolumn{2}{l}{Tax payer}                          \\ \cline{2-3} 
\multicolumn{1}{l|}{\multirow{2}{*}{Auditor}} & \multicolumn{1}{l|}{2,0} & \multicolumn{1}{l|}{4,-20} \\ \cline{2-3} 
\multicolumn{1}{l|}{}                          & \multicolumn{1}{l|}{4,0} & \multicolumn{1}{l|}{0,4} \\ \cline{2-3} 
\end{tabular}%

}}
\end{table}
\subsubsection{Strategic Resource Allocation}\label{strategicres}

Problems of resource allocation and fair division have received considerable research attention from a computational perspective, as they are a critical consideration in multi-agent systems. In a general resource allocation problem, there are resources that we want to distribute among agents with different valuations over the resources (sometimes bundles of resources). Fair division is a type of social choice problem involving a group of agents where each has individual preferences that are usually represented by a utility function. The goal of fair division is the allocation of resources in a fair and equitable manner. A fundamental question underlying the fair division problem is how to define fairness criteria in the first place. Much research in economics, especially social choice theory has addressed this topic. Classical fairness definitions include competitive equilibrium \cite{Arrow} and social Nash welfare \cite{socialnashwelfare}. Many fairness notions for indivisible goods have been considered; however, many of these notions are hard to compute \cite{Fairdivistra,AAAI,approxsocialnash,Branzei}. Even further, almost all of the well-known works are limited to non-strategic settings.

Strategic fair division is a branch of fair division in which participants may act uncooperatively in order to maximize their own utility.  In particular, the players may hide their true preferences, rather than playing sincerely according to their true preferences.  In the presence of participants who have strategic behavior, it is essential to have a suitable fair allocation with appropriate fairness criteria in place to allocate resources in a fair and equitable manner. One branch of fair division is related to game theory and studies the equilibrium in games resulting from fair division algorithms \cite{Brnzei2016,Branzei,Gamesdivis}.

An envy-free allocation is an allocation in which each player receives a share that is, in their eyes, at least as good as the share received by any other agent \cite{puzzle}. There are several different variations of envy-freeness \cite{superenvy,barman2019fair,Super2}. In  the strategic setting, we define a variation of envy-freeness in the following:

\begin{definition}
    Suppose that we have $k$ players and $n$ items. The game $\mathcal{RV}$ is defined as follows:
$\forall i\in[k]~\text{and}~ \forall t\in [n]$, $U_i(t,x_{-i})=\alpha V_i(t)+ \beta R_i(x_{-i})$,
where $\alpha$ and $\beta$ are  constants. $V_i(t)$ indicates the value of item $t$ for player $i$ and $R_i(x_{-i})$ is a function that is computable in constant time that shows how a player can interpret the allocation of items (pure strategies) $x_{-i}=(t_1,\dots,t_{i-1},t_{i+1},\dots,t_n)$ to the other players. 
\end{definition}

\begin{definition}
  An allocation is an \emph{$(\delta,\epsilon)$-item-wise envy-free}  allocation if  each player receives the items that are in the support of a $\epsilon$-partition Nash equilibrium $x^*$ of game $\mathcal{RV}$ with the following properties:
    \begin{itemize}
    \item Each player $i\in [k]$ has a support of size $\frac{n}{k}$.
      \item For each player $i\in [k]$, we have $||x^*_i-\frac{k}{n}||_1\leq \delta$.
    \end{itemize}

\end{definition}

\begin{remark}
   It is easy to observe that in a \emph{$(0,0)$-item-wise 
   envy-free}  allocation, no player would prefer to exchange any item allocated to another player for any of the items in their own allocation. Note that in an envy-free allocation, each player is only concerned with the value of the items.
\end{remark}

\begin{proposition}\label{itemwise}
 The decision problem of whether there exists \emph{$(\delta,\epsilon)$-item-wise envy-free} allocation is NP-complete for some $\delta$ and $\epsilon$ polynomially bounded in the size of the game.
\end{proposition}
\begin{proof}
    This immediately follows by the hardness of partition Nash.
\end{proof}

\begin{remark}
     Another branch of fair division considers truthful resource allocation mechanisms, in which agents are incentivized to reveal their true valuations for resources. We do not address truthful mechanisms in this paper.
\end{remark}

\subsection{Main Theorems}
A number of the problems we consider depend on parameters besides the additive approximation term. These parameters depend on $n$, which is the size of the strategy space $S$. We say that such a parameter $\alpha$ is \emph{poly-bounded} if there is some $k>0$ such that $\alpha(n)=O(\frac{1}{n^k})$. For Nash equilibrium with constraints, we have the following results:

\begin{theorem}\label{deltafarhardness} For bi-matrix games, the following problems are NP-complete under POLY-approx:
  \begin{enumerate}[(a)]
       %\item  $\delta$-Far Nash, for any $\delta$ polynomial in $|S|$, 
       \item  $\delta$-Far Nash, for any  poly-bounded $\delta$
       \item Disjoint Nash
       \item Partition Nash
        %\item More than one Nash equilibrium with a statistical difference of $\delta$ for some $\delta$ polynomial in $|S|$.
        \item More than one Nash equilibrium with a statistical difference of $\delta$ for some poly-bounded $\delta$
       %\item $\theta$-Major Nash equilibrium  (for some $\theta$ polynomial in $|S|$).
       \item $\theta$-Major Nash equilibrium for some poly-bounded $\theta$ 
   \end{enumerate}
\end{theorem}
%Here $S$ denotes the strategy set of the given bi-matrix game $\mathcal{G}$, and $\delta,\theta$ are represented in unary (i.e., $\frac{1}{k}$ is represented by $1^k$.) 
For the EXACT version, it is possible to extend the approach of  Gilboa and Zemel \cite{Gilboa} to show that disjoint Nash is NP-hard.

We may consider the question of whether there is a threshold approximation error below which $\delta$-far Nash becomes a total problem. Despite the absence of a definitive answer to this question, we provide a bound within which a $\delta$-far Nash equilibrium is guaranteed to exist in the following theorem:

\begin{theorem}\label{sdfarmin}
	For any game $\mathcal{G}$, there exists a $\delta(n) \geq \frac{1}{n}$ such that at least one $(\delta,4\delta$)-far Nash equilibrium for $\mathcal{G}$ exists. Furthermore, this problem is in PPAD.
\end{theorem}

 The main complexity results related to the constrained equilibrium version of these problems are  the following:

%\koosha{Should we say this? instead}

\begin{proposition}\label{polyconst}
 For any game $\mathcal{G}$, the problem of finding a constrained disjoint equilibrium under EXP-approx is computable in polynomial time.
\end{proposition}

%\koosha{Changed tiny bit}
The following theorems are all with respect to bi-matrix games and under POLY-approx.

\begin{theorem}\label{farppadhard}

 There is a poly-bounded $\delta$ for which the problem of finding a $\delta$-far constrained equilibrium is PPAD-hard.    
 \end{theorem}

 \begin{theorem}\label{farinppad}
There is a poly-bounded $\delta$ such that the problem of finding a $\delta$-far constrained equilibrium for games whose diagonally modified version does not admit a fully mixed Nash equilibrium (see Definition \ref{refdmg}) is in $PPAD$.
\end{theorem}
 
\begin{remark}
    The previous assumption can be easily checked by using part of a well-known  \emph{support enumeration} algorithm. For more details, see Appendix~\ref{supportenusection}.
\end{remark}

 \begin{theorem}\label{disjointrestricted}
  For some poly-bounded $\theta$,  the problem of deciding whether there exists a $\theta$-restricted disjoint equilibrium is NP-complete.
 \end{theorem}

\begin{theorem}\label{farrestrict2}
  For some poly-bounded $\theta$ and any poly-bounded $\delta$, the problem of deciding whether there exists a $(\theta,\delta)$-restricted far equilibrium is NP-hard.
 \end{theorem}

\begin{remark}
    The stated theorems admit a number of hardness results of other variations of Nash equilibrium including additional information about the parameters. We have not stated them here in the interest of simplifying the presentation.\end{remark}

\subsection{Technical Overview}
The NP-hardness proofs of the problems stated in this paper are inspired by the generic proof  provided by \cite{Conitzer} for NP-hardness of deciding the existence of a Nash equilibrium with constraints. This generic reduction (from SAT which is an NP-complete problem) improved the NP-completeness results for Nash equilibrium with certain constraints given in \cite{Gilboa} (which are only concerned about the EXACT versions).  The proof of inapproximability in \cite{Conitzer} does not apply to the form of approximation that we consider in this paper (the additive approximation error introduced in \cite{ComplexityNASH}). However, a reduction that is provided by Schoenebeck and Vadhan \cite{vadhan} (which modifies the proof provided in \cite{Conitzer}) shows that the problem of finding a Nash equilibrium with a certain guaranteed payoff for all players even under POLY-approx is NP-complete. Throughout the paper, we introduce and use a standard \emph{duplication} technique (see Example \ref{changelabellemma}) in which we make two copies of some of the strategies where each player will have its own \emph{associated} strategy while the other copy (which we call \emph{unassociated} strategy) will be a strictly dominated strategy. We show that, under any approximation scheme that we reviewed, this technique can connect present hardness results on Nash equilibrium to our strategic constraints if applied carefully.  

The NP-hardness results of this paper are derived  by combining the techniques that we introduce and ideas given in both reductions given in \cite{Conitzer} (Corollary $6$), \cite{vadhan} (Theorem $8.6$). For clarity, we break the reduction \cite{vadhan} down into a series of lemmas each of which will be used frequently in different proofs with some modest modifications. We begin with the hardness of disjoint Nash that directly applies (an adapted version of) these lemmas to a game constructed by modifying the game provided \cite{Conitzer} along with considering two other possible cases that need to be taken into account. We also introduce another standard technique that bounds the approximation error under strategy modification (see Lemma \ref{blowup}). The hardness of partition Nash equilibrium and the existence of approximate far Nash equilibrium in certain cases also follows by using this lemma along with some other techniques (see Theorem \ref{sdfarmin} and Proposition \ref{expartiiton}). Through the clever and nontrivial expansion of the duplication technique, we are able to narrow down the distance of far Nash equilibrium and prove its hardness (see Section \ref{farproofhardsection}). For simplicity, we break the proof down into two sub-cases where the statistical distance can be either a small or big number (see \ref{Hdelta} and \ref{deltahardfinal}.)

Towards exploring the computational complexity of constrained far equilibrium, we utilize another technique that penalizes the diagonals of the payoff matrices of the players. By using this approach, we demonstrate that Nash equilibrium with certain properties can be related to the concept of constrained equilibrium when farness-related constraints are considered. The hardness of restricted equilibrium and Nash equilibrium without minor probabilities will be shown by embedding a sub-game similar to \emph{rock-paper-scissors} into the previously mentioned games that were used in reductions and also combining all of the discussed techniques (see Theorem \ref{disjointrestricted} and \ref{farrestrict2}.)

\section{Proofs of Main Theorems}

\subsection{Computational Complexity of Far Nash and Theorem \ref{deltafarhardness}}
It is obvious that all of the stated problems are in NP. Our proof for the NP-hardness results stated in Theorem \ref{deltafarhardness} begins with proving that POLY-approx disjoint Nash (Theorem \ref{deltafarhardness} part (b)) is NP-hard (shown in Proposition \ref{modifedconitzer}), from which it follows that far Nash (for $\delta=1$) is NP-hard as well.  Next, the hardness of POLY-approx partition Nash equilibrium (Theorem \ref{deltafarhardness} part (b)) can be derived by a modest modification of Proposition \ref{modifedconitzer} (see Proposition \ref{expartiiton}). To prove Theorem \ref{deltafarhardness} part (a), we show that the hardness can be extended to any $\delta$ polynomial in the size of the strategies in the game (see Section \ref{farproofhardsection}).  The hardness of the problem of whether a game has more than one Nash equilibrium with a fixed statistical distance (Theorem \ref{deltafarhardness} part (d)) is proved in Corollary \ref{uniqueanddisjoint}.   Finally, we finish the proof of this theorem by showing the hardness of the POLY-approx major Nash (Theorem \ref{deltafarhardness} part (e)) in Corollary \ref{majornashhard}.

Throughout the paper, we use two standard techniques in multiple situations. To derive hardness results for disjoint Nash through standard NP-hard reductions from well-known NP-hard problems, we need to ensure that the required game solutions have disjoint support. This is done by duplicating some of the strategies so that the players can only play strategies from their \emph{associated} sets. We illustrate this technique with the following example:

\begin{exmp} \label{changelabellemma}
	The problem of finding an EXACT Nash equilibrium is poly-time reducible to finding an EXACT disjoint Nash equilibrium.
\end{exmp}
 \begin{proof} Given a bi-matrix game $\mathcal{G}$ with $S=[n]$ strategies, we create game $\mathcal{G^{\prime}}$ and prove that  a Nash equilibrium in $\mathcal{G}$ will form a disjoint Nash equilibrium in $\mathcal{G^{\prime}}$.  Let $\sigma$ be a value that is strictly less than the minimum payoff for either player in $\mathcal{G}$. In $\mathcal{G^{\prime}}$, all players  have the strategy space $S^{\prime}=S_1 \cup S_2$ where $S_1=[n]$ and $S_2=n+[n]$ are just copies of $S=[n]$. The new payoff matrix will force each player $i$ to only play from set $S_i$. For any $(i,j)$ from $S^{\prime}\times S^{\prime}$, the utility matrix for each player is defined as follows:
\[
	U^{\prime}_{1}(i,j) = 
	{
	\begin{cases}
		U_{1}(i,j-n), & \text{if } i\in S_1\\
	\sigma & \text{o.w }
	\end{cases}
	} 
\]
\[
	U^{\prime}_{2}(i,j) = 
	{
	\begin{cases}
		U_{2}(i,j-n), & \text{if } j\in S_2\\
	\sigma & \text{o.w }
	\end{cases}
	} 
\]
	
	     Suppose that $(x^*,y^*)$ is an EXACT Nash equilibrium for $\mathcal{G}$. If each player $k$ plays the same distribution on members of $S_k$ that was played on $S$ in $(x^*,y^*)$, we have an EXACT disjoint Nash equilibrium for $\mathcal{G^{\prime}}$, by the construction and strict domination of $\sigma$.  Now suppose $(x^{\prime},y^{\prime})$ is a Nash equilibrium with disjoint supports in $\mathcal{G^{\prime}}$. We easily can map these distributions to a Nash equilibrium $(x^*,y^*)$ of $\mathcal{G}$ by defining, for $i \in S$,
$ x^*_{i}=x^{\prime}_{i}$ and $y^*_{i}=y^{\prime}_{i+n}$.

\end{proof}

\begin{remark}
The preceding technique can only be applied to the EXACT versions of our problems and needs to be adapted to work with the approximate version.
\end{remark}

\begin{lemma}\label{blowup}
	 Suppose that $(x^*,y^*)$ is a Nash equilibrium in a bi-matrix game whose utility matrix entries are in the range $[\alpha,\beta]$. For each $i$, we define $x^-_i$ to be the same as $x^*_i$ except we take $\epsilon_i\leq x^*_i$ from the weight of strategy $i$  and redistribute it to some strategies in $[n]-\{i\}$. Assuming $\sum_{i=1}^{n} \epsilon_i\leq\epsilon$, we have the following:
	
	\begin{enumerate}
	    \item $(x^-,y^*)$ and $(x^*,y^-)$ are $2\epsilon (\beta-\alpha)$-Nash equilibria.
	    \item 	$|U_1(x^*,y^*)-U_1(x^-,y^-)|\leq 2\epsilon (\beta-\alpha)$
	     \item 	$|U_2(x^*,y^*)-U_2(x^-,y^-)|\leq 2\epsilon (\beta-\alpha)$
	\end{enumerate}

\end{lemma}
\begin{proof}
    See Appendix A. We assume $\alpha=0$, $\beta=1$ unless stated.
\end{proof}

Before proving the NP-hardness of disjoint Nash, we recall the satisfiability problem for 3CNF formulas and some associated terminology.

\begin{definition}
	 A \emph{Boolean formula $\phi$  in CNF} (conjunctive normal form) is specified by a set $V$  of \emph{variables} (with $|V|=n$), a set of $L$ of \emph{literals} consisting of variables and their negations, and a set $C$ of clauses, where each clause is a set of literals. A \emph{3CNF formula} is a CNF formula in which each clause has exactly 3 literals. 3CNFSAT is the problem of deciding whether there is a satisfying assignment for a 3CNF formula $\phi$ (i.e. a setting of the variables to \emph{true} or \emph{false} under which $\phi$ evaluate to \emph{true}.)
  \end{definition}

To show the hardness of POLY-approx disjoint Nash, we give a poly-time mapping of any 3CNF $\phi$ to a two-player game $\mathcal{G}(\phi,\epsilon)$ and prove the following proposition.
\begin{proposition} \label{modifedconitzer}
	Given an instance of \text{3CNF} $\phi$ with $n$ variables, there is a $\epsilon$-disjoint Nash equilibrium (where $\epsilon=\frac{1}{2n^3}$) in $\mathcal{G}(\phi,\epsilon)$ iff  $\phi$ is satisfiable.
\end{proposition}

The game $\mathcal{G}(\phi,\epsilon)$ is a modification of the game given in \cite{vadhan}, which we refer to as $\mathcal{SV}(\phi,\epsilon)$.\footnote{The proof of hardness of deciding whether $\mathcal{SV}(\phi,\epsilon)$ has a $\epsilon$-Nash with a guaranteed payoff $n-1-\epsilon$ is given in Appendix B.} 
$\mathcal{G(\phi,\epsilon)}$ is a  non-symmetric $2$-player normal form game given  as follows: Let $S \equiv S_{1}=S_{2}=L_1 \cup L_2 \cup V \cup C \cup\{f\}$ be the strategy set for both players, where $L_1$ and $L_2$ are copies of $L$. Let  $\mathrm{v}: L \rightarrow V$ be the function that gives the variable corresponding to a literal, e.g. $\mathrm{v}\left(x_{1}\right)=\mathrm{v}\left(-x_{1}\right)=x_{1}$. For example, if $x_1$ is a variable, $x_1$ and $-x_1$ are literals that are \textbf{representatives} of variables $x_1$ being true or false respectively. Also, let  $\mathrm{g}:L_1 \cup L_2 \rightarrow L$ be the function (which we call the \emph{unlabelling function}) that maps copies  $L_1$ and $L_2$ of $L$ to $L$.  The utility matrices are defined as follows:

\begin{enumerate}

	\item $u_{1}\left(l^{1}, l^{2}\right)=u_{2}\left(l^{1}, l^{2}\right)=n-1$ for all $l^{1} \in L_1 ,~ l^{2} \in L_2$ with $\mathrm{g}(l^{1} )\neq-\mathrm{g}(l^{2})$;
	
	\item $u_{1}(l^1,-l^2)=u_{2}(-l^1, l^2)=n-4$ for all $l^1 \in L^1, ~, l^2 \in L^2 ~\&~ \mathrm{g}(l^1)=\mathrm{g}(l^2)$;
	
	\item $u_{1}(l^1, x)=u_{2}(x, l^2)=n-4$ for all $l^{1} \in L_1 ,~ l^{2} \in L_2, ~ x \in V \cup C$;
	
	\item $u_{1}(v, l^2)=u_{2}(l^1, v)=n$ for all $v \in V,~ l^1 \in L_1, ~ l^2 \in L_2 ~\&~ \mathrm{g}(l^1)=\mathrm{g}(l^2)$ with $\mathrm{v}(\mathrm{g}(l^1)) \neq v$;
	
	\item $u_{1}(v, l^2)=u_{2}(l^1, v)=0$ for all $v \in V,~l^1 \in L_1, ~ l^2 \in L_2 ~\&~ \mathrm{g}(l^1)=\mathrm{g}(l^2)$ with $\mathrm{v}(\mathrm{g}(l^1))=v$;
	
	\item $u_{1}(v, x)=u_{2}(x, v)=n-4$ for all $v \in V,~ x \in V \cup C$;
	
	\item $u_{1}(c, l^2)=u_{2}(l^1, c)=n$ for all $c \in C,~ l^1 \in L_1, ~  l^2 \in L_2$ with $\mathrm{g}(l^1)=\mathrm{g}(l^2) \notin c$;
	
	\item $u_{1}(c, l^2)=u_{2}(l^1, c)=0$ for all $c \in C,~ l^1 \in L_1, ~ l^2 \in L_2$  with $\mathrm{g}(l^1)=\mathrm{g}(l^2) \in c$;
	
	\item $u_{1}(c, x)=u_{2}(x, c)=n-4$ for all $c \in C,~ x \in V \cup C$;
	
	\item $u_{1}(x, f)=u_{2}(f, x)=0$ for all $x \in S-\{f\}$;
	
	\item $u_{1}(f, f)=u_{2}(f, f)=2n$;
	
	\item $u_{1}(f, x)=u_{2}(x, f)=n-1$ for all $x \in S-\{f\}$.
	
\end{enumerate}

The following cases also need to be defined as they can possibly appear when players do not play from their associated literal set. We define them in such a way that they become strictly dominated strategies:

\begin{enumerate}
 \setcounter{enumi}{12}
\item $
u_{1}(l^2,*)=u_{2}(*,l^1)=-2n \thickspace \mbox{for all} \thickspace *\in S,~  l^2 \in L_2 ~\text{and}~ l^1 \in L_1;
$
\item $
u_{2}(l^2,*)=u_{1}(*,l^1)=0 \thickspace \mbox{for all} \thickspace *\in S-\{f\},~  l^2 \in L_2 ~\text{and}~ l^1 \in L_1.
$
\end{enumerate}
\begin{remark}\label{badrules}
Note that in \cite{Conitzer}, $U_i(f,f)$ is an arbitrary positive number. 
\end{remark}

\subsubsection{Essential Lemmas for Proposition \ref{modifedconitzer}}

Before proving  Proposition \ref{modifedconitzer}, we need several lemmas.

\begin{lemma}\label{smallnonassociate}
In any $\epsilon$-Nash equilibrium of $\mathcal{G(\phi,\epsilon)}$, the row player cannot play literals from $L_2$ with  probability greater than $\frac{\epsilon}{2n}$. A symmetric argument also applies to the column player and all other strategies that have a payoff of $-2n$.         
\end{lemma}

\begin{proof}
    If one player plays a literal not from its  associate literal set, they will get a payoff of $-2n$.  Thus, in any $\epsilon$-Nash equilibrium, these strictly dominated strategies  can be played with a probability of at most $\frac{\epsilon}{2n}$, as playing such a strategy with probability greater than $\frac{\epsilon}{2n}$ will cause the player to lose at least $\epsilon$ compared to other strategies which violates the definition of $\epsilon$-best response.
\end{proof}

\begin{lemma}
	\label{maxsw2}
    In any $\epsilon$-Nash equilibrium of $\mathcal{G}(\phi,\epsilon)$ if neither player plays $f$ with a positive probability, literals that are not associated with the players, as well as clauses and variables,  can be played with a probability of at most $\epsilon$.
\end{lemma}
\begin{proof}
    The proof is similar to Lemma \ref{maxsw}.
\end{proof}

Consequently, strictly dominated strategies, variables, and clauses cannot be played with a combined probability greater than $\epsilon$. The next lemma will show the fact that if one of the players cannot meet the guaranteed payoff $n-1-\epsilon$, an $\epsilon$-approximate disjoint Nash equilibrium cannot be obtained.

\begin{lemma}\label{gurantee}
    In any $\epsilon$-Nash equilibrium in $\mathcal{G}(\phi,\epsilon)$, all players have a guaranteed payoff of at least $n-1-\epsilon$.
\end{lemma}
\begin{proof}
Suppose that one of the players does not have a guaranteed payoff of at least $n-1-\epsilon$. This player can deviate from the assumed strategy and play $f$ to get $n-1$ no matter what the other player does. Consequently, this strategy is not a $\epsilon$-best response to the opponent player. 
\end{proof}

\begin{lemma}
    Suppose that the formula $\phi$ is satisfiable. The game $\mathcal{G(\phi,\epsilon)}$, has at least one EXACT disjoint Nash equilibrium. 
\end{lemma}
\begin{proof}
  An argument that is similar to  Theorem~\ref{SVgame} can be applied to $\mathcal{G(\phi,\epsilon)}$ to show that there exists an EXACT disjoint Nash equilibrium. The only difference here is the duplicated literal sets and the unlabelling function. In other words, the strategy where each player randomizes uniformly over $n$ out of $2n$ (based on a satisfying assignment) of their associated literals in $L_1$ and $L_2$ respectively, is an EXACT disjoint Nash equilibrium (where the expected payoff to each player is $n-1$.)  
\end{proof}

\begin{lemma}\label{sum1ndisjoint}
      Suppose that $\phi$ is not satisfiable. If neither player plays $f$ with a positive probability, then for any $l^1\in L_1$, the probability that the row player plays $l^1\in L_1$ or $-l^1\in L_1 $ is at least  $\frac{1}{n}-2 \epsilon$. 
\end{lemma}

\begin{proof}
    
  Suppose this is not the case. Assume that the probability is less than $ \frac{1}{n} -\epsilon-\frac{2 \epsilon}{n} \geq \frac{1}{n}-2 \epsilon $. Let $l^2\in L_2$ be a literal such that $\mathrm{g}(l^2)=\mathrm{g}(l^1)$.
	  Recall that in Lemma \ref{maxsw2},  both players play from their associated literal sets with a probability of at least $1-\epsilon$ while all other strategies can be played with a probability of at most $\epsilon$. Then, the expected payoff for the column player, when playing $\mathrm{v}(\mathrm{g}(l^2))$ will be at least $n-1+2\epsilon$ by Lemma \ref{lornegativel} where the only modification is that we use Lemma \ref{maxsw2} instead of Lemma \ref{maxsw}.  Since the maximum social welfare is $2n-2$, the other (row) player fails to meet the guarantee $n-1-\epsilon$ and will deviate to $f$. 
\end{proof}
\begin{remark}
    A similar argument can be applied to the column player as well.
\end{remark}

\begin{lemma}\label{nofonegreater}
      Suppose that $\phi$ is not satisfiable. If neither player plays $f$ with a positive probability, then for each player and literal $l$ from their associated set, either $l$ or $-l$ is played with probability $\geq$ $\frac{1}{ n} -2\epsilon-\frac{1}{ n^{2}}$ while the other is played with probability less than $\frac{1}{ n^{2}}$.
\end{lemma}

\begin{proof}
    	 The proof is exactly the same as Lemma \ref{bothliterals} except each player will play from their respective  literal set with a high probability.
\end{proof}

\begin{lemma}\label{gurantee2}
     Suppose that $\phi$ is not satisfiable. If neither player plays $f$ with a positive probability, at least one player cannot guarantee a payoff of $n-1-\epsilon$.
\end{lemma}

\begin{proof}

		We prove that in any $\epsilon$-Nash equilibrium in the constructed game except for the ones that have $f$ in their support, there still exists a one-to-one correspondence between literals and truth assignments.  We know that for each player, the player will play one of $l$ or $-l$ (from their associated literals) with a probability of at least $1-\frac{1}{n^2}-2\epsilon$ while the other is played with a probability of at most $\frac{1}{n^2}$. We can consider the variables to be true if $l^1$ from $L_1$ and $l^2$ from $L_2$ are played with a higher probability compared to  $-l^1$ and $-l^2$ for each player respectively (note that $\mathrm{g}(l^1)=\mathrm{g}(l^2)=l$). If an assignment does not satisfy the formula, by changing the strategy to a clause that is not satisfiable, one of the players will receive at least $\left(1-\epsilon-\frac{3}{n^{2}} \right) n>n-1+2 \epsilon$ by Proposition \ref{SVgame}. So, the other player cannot attain $n-1-\epsilon$.

\end{proof}

\subsubsection{Proof of Proposition \ref{modifedconitzer}}

\begin{proof}

	    We showed that if $\phi$ is satisfiable, an EXACT disjoint Nash equilibrium (also an approximate solution) exists. We also showed that if $\phi$ is not satisfiable, if neither player plays $f$ with a positive probability, at least one player cannot guarantee a payoff of $n-1-\epsilon$. In conclusion, by Lemma~\ref{gurantee}, one player will have to play $f$ and we show that this forces the other player to use $f$  with a positive probability.

	Now, assume that one of the players plays $f$ with probability $\alpha$ while the other player does not play $f$. There are two possible cases for $\alpha$ where the first case is that $\alpha<\frac{\epsilon}{n}$ and the second case is  $\alpha\geq\frac{\epsilon}{n}$. We show that for either of the cases, an $\epsilon$-approximate disjoint Nash equilibrium cannot be generated in $\mathcal{G}(\phi,\epsilon)$ in the following lemma.
		\end{proof}

  \begin{lemma}\label{fmodifiedconitzer}
      Let  $\epsilon=\frac{1}{2n^3}$ where $n$ denotes the number of the variables in $\phi$. If $\phi$ is unsatisfiable, $\mathcal{G}(\phi,\epsilon)$ has no $\epsilon$-approximate Nash equilibrium in which exactly of the players plays $f$ with a positive probability.
  \end{lemma}
	
    \begin{proof}

     Suppose that the row player plays $f$ with probability $\alpha$. It is easy to observe that no $\epsilon$-Nash equilibrium with a payoff less than $n-1-\epsilon$ exists for both players similar to Lemma \ref{gurantee}.  The column player can actually guarantee a better payoff of $[\alpha\cdot (2n)+(1-\alpha)(n-1))]-\epsilon$ by deviating to $f$. As argued above, if one of the players does not play $f$, the maximum possible social welfare is at most $2n-2$. So if one of the players can achieve a payoff greater than $n-1+2\epsilon$, the other player will get less than $n-1-\epsilon$. If one of the players plays $f$ with a positive probability, variables, clauses, and literals that are not associated with players, can be played with a probability of at most $\epsilon$ (we will use this bound when determining payoffs later). This follows as in Lemma~\ref{maxsw2} since the only modification in this setting is that the column player has a better guaranteed expected payoff.

	Now assume that $(x^*,y^*)$ is an approximate disjoint Nash equilibrium where the row player plays $f$ with probability $0<\alpha<\frac{\epsilon}{n}$ and the column player plays $f$ with probability $0$. We show (similarly to Lemma \ref{lornegativel}) that the probability that the row player plays $l^1$ or $-l^1$ is at least $\frac{1}{n}-2\epsilon$: Suppose that the probability is less than $\frac{1}{n}-2\epsilon$. The expected payoff for the column player playing $\mathrm{v}(\mathrm{g}(l^2))$ (we know $\mathrm{g}(l^1)=\mathrm{g}(l^2)$) with  probability $1$ is at least $n-1+2\epsilon$, computed as follows:
	
	\begin{itemize}
		\item 	When the row player plays $l^1$ or $-l^1$ the payoff is zero.
		
		\item When the row player plays an associated literal in $L_1$ other than  $l^1$ or $-l^1$, the payoff is
		\[(1-\epsilon-(\frac{1}{n}-2\epsilon)-\alpha) n = (1+\epsilon-\frac{1}{n}-\alpha)n\geq n+\epsilon n -1 -\epsilon=n-1+\epsilon(n-1) \]
		
		\item When the row player does not play a literal the payoff is at least $\epsilon \cdot 0$.

		\item When the row player plays $f$, the portion of the payoff is $\alpha\cdot0$. 
		
	\end{itemize}

	The summation will be at least $n-1+2\epsilon$ (for $n>2$) which indicates the row player cannot meet the guarantee $n-1-\epsilon$. In conclusion, it is better for the row player to play $f$ with a greater probability namely $1$, not $\alpha$. For any $l^2\in L_2$, the probability that the column player plays either $l^2$ or $-l^2$ is at least $\frac{1}{n}-2\epsilon$ by Lemma \ref{sum1ndisjoint}. This is because the column player does not play $f$. So, for each player, the probability of playing either $l$ or $-l$ from their associated literal set is at least $\frac{1}{n} -2\epsilon$. This implies that one of $l^1\in L_1$ or $-l^1$ (for any $l^1 \in L_1$) is played with  probability greater than $\frac{1}{ n} -2\epsilon-\frac{1}{ n^{2}}$ while the other is played with  probability smaller than $\frac{1}{n^2}$ (see Lemma \ref{bothliterals} and \ref{nofonegreater}). The preceding holds for the column player and the associated literal as well. Using an argument similar to those used in Lemma \ref{gurantee2} and Proposition \ref{SVgame}, we conclude that the row player can change its strategy to play an unsatisfied clause (with a high probability $1-\alpha$) and receive at least $(1-\epsilon-\frac{3}{n^{2}}-\alpha )n>n-1+2 \epsilon$. Since the maximum social welfare is $2n-2$, it is better for the column player to play $f$ and obtain a payoff of $n-1$ instead.

 Now assume the row player plays $f$ with probability $\alpha > \frac{\epsilon}{n}$ while the column player plays $f$ with probability zero. Recall that the column player can only play strategies other than its associated literals with a probability of at most $\epsilon$. If the column player deviates to  $f$ with probability $1$, they will increase their payoff by at least $\epsilon$  which violates the definition of $\epsilon$-best response:
	\[[2\alpha\cdot n+(1-\alpha)(n-1)]-[(1-\epsilon-\alpha)(n-1)+\epsilon \cdot n +\alpha\cdot0]=2\alpha\cdot n -\epsilon>\epsilon\]
	\end{proof}

\begin{remark}
	The minimum and the maximum payoff of the game $\mathcal{G(\phi,\epsilon)}$ are $-2n$ and $2n$ respectively. We can scale the payoffs of this game to be in the standard range $[0,1]$. 
\end{remark}

The following corollary establishes the hardness of deciding whether a game has more than one equilibrium with  statistical distance at least $\delta=2\epsilon$ between the equilibria of the game (Theorem \ref{deltafarhardness} part (d)).  

\begin{corollary}\label{uniqueanddisjoint}
    When the formula $\phi$ is unsatisfiable, both players will have to play $f$ with  probability greater than $1-\epsilon$ in any equilibria of the game $\mathcal{G(\phi,\epsilon)}$. 
\end{corollary}
\begin{proof}
    We can proceed as in the proof of Proposition \ref{modifedconitzer}. Assume that the row player plays $f$ with  probability  $\alpha_1$ and the column player plays $f$ with $\alpha_2$. It is easy to see that one of $\alpha_1$ or $\alpha_2$ must be greater than $\frac{\epsilon}{n}$. If both of these probabilities are smaller than $\frac{\epsilon}{n}$, one of the players cannot reach their guaranteed payoff, by a modification of Lemma \ref{fmodifiedconitzer} (both players play $f$ with a small probability). In conclusion, we can assume that $min(\alpha_1,\alpha_2)\geq \frac{\epsilon}{n}$. It is easy to show that both players have a guaranteed payoff of at least $[min(\alpha_1,\alpha_2)\cdot (2n)+(1-min(\alpha_1,\alpha_2)(n-1))]-\epsilon$ by just playing the pure strategy $f$. The players can only achieve a payoff greater than their approximate guaranteed payoff only if they play $f$ with  probability greater than $1-\epsilon$. If one player plays $f$ with probability smaller than $1-\epsilon$ (we know that must be greater than $\frac{\epsilon}{n}$), the player loses $2\min(\alpha_1,\alpha_2)\cdot n-\epsilon$ which is greater than $\epsilon$.
\end{proof}

\begin{remark}
    Note that when the formula is satisfiable, an EXACT disjoint Nash equilibrium exists in addition to the Nash equilibrium with the strategy profile $(f,f)$.
\end{remark}

\subsubsection{Hardness of Partition Nash}

 While proving EXACT partition Nash is NP-hard appears to be quite challenging, an NP-hardness result for $2$-player POLY-approx partition Nash equilibrium can easily be derived from the NP-hardness result that we have for disjoint Nash equilibrium and Lemma \ref{blowup}.

\begin{proposition}
    \label{expartiiton}
   Given an instance of 3CNF $\phi$ with $n$ variables, the game $\mathcal{G}(\phi,\epsilon)$ has a $\epsilon$-partition Nash equilibrium (for $\epsilon=\frac{1}{2n^3}$) iff the formula is satisfiable.
\end{proposition}
\begin{proof}
       See Appendix B.
\end{proof}
\subsubsection{Hardness of Far Nash and Theorem \ref{deltafarhardness}}\label{farproofhardsection}
In this section, we complete the proof of Theorem \ref{deltafarhardness}. We prove that for any $\delta$ and some $\epsilon$  polynomially bounded in the size of the strategies in a game, the problem of deciding whether there exists a $(\delta,\epsilon)$-far Nash remains hard.

Given a formula $\phi$ with $n$ variables, we generate the game $\mathcal{H}(\phi,\delta,\epsilon)$, which is similar to $\mathcal{G}(\phi,\epsilon)$ with the exception that not all literals are duplicated, depending on the given value $\delta$ (see Proposition \ref{Hdelta} and \ref{deltahardfinal}). The strategy space for both players is $S^{H_{\delta}} \equiv S^{H_{\delta}}_1=S^{H_{\delta}}_2=L^{\prime}\cup L^{\prime}_1 \cup L^{\prime}_2 \cup V \cup C \cup\{f\}$. We divide the  original literals ($L$ from the formula $\phi$)  into three sets. The set $L^{\prime}$ will contain the literals from $L$ that are not duplicated, while  $L^{\prime}_1$ and $L^{\prime}_2$ contain the duplicated literals, associated respectively with the row and the column player. The function $\mathrm{g}^{\prime}: L^{\prime}\cup L^{\prime}_1 \cup L^{\prime}_2 \rightarrow L$ can be defined similarly to $\mathrm{g}$  except that it maps elements of $L^{\prime}\subset L$ to themselves. The rules for literals in $L^{\prime}$ are similar to rules in $\mathcal{SV}(\phi,\epsilon)$ and the ones are in $ L^{\prime}_1 \cup L^{\prime}_2$  similar  to those in $\mathcal{G}(\phi,\epsilon)$.

\begin{proposition}\label{Hdelta}
 Given any 3CNF formula $\phi$ with $n$ variables, for any $\delta>\frac{1}{n}-\frac{1}{n^2}-\frac{1}{n^3}$, there exists a game $\mathcal{H}(\phi,\delta,\epsilon)$ such that for some $\epsilon$ polynomially bounded in $n$, there exists a $(\delta,\epsilon)$-far Nash equilibrium in $\mathcal{H}(\phi,\delta,\epsilon)$ iff $\phi$ is satisfiable.
\end{proposition}

\begin{proof}
    The proof follows by combining Proposition \ref{modifedconitzer} and Proposition \ref{SVgame}.   Recall that by Lemma \ref{nofonegreater}, \ref{fmodifiedconitzer} and \ref{bothliterals}, for any literal $l\in L^\prime$  (and also the associated copies $l_1\in L_1^\prime$ and $l_2\in L_1^\prime$), either the literal or the negation of the literal $-l$  will be played with a probability of at least $\frac{1}{n}-2\epsilon-\frac{1}{n^2}$ when both players do not play $f$ with a positive probability greater than $\frac{\epsilon}{n}$. 
    
    To generate $\mathcal{H}(\phi,\delta,\epsilon)$ for the given $\delta$, we determine $i$ such that $\delta=\frac{1}{2}(2i(\frac{1}{n}-2\epsilon-\frac{1}{n^2}))$ and select $\lfloor i \rfloor$ random literals of the original formula and generate the sets $L_1^\prime$ and $L_2^\prime$. The rest of the literals will form $L^\prime$ (will not be duplicated). Note that for $\delta^\prime \leq \delta$, NP-hardness of $(\delta^\prime,\epsilon)$-far Nash implies the hardness of $(\delta,\epsilon)$-far Nash. An argument similar  to Corollary \ref{uniqueanddisjoint} implies that if the given formula is unsatisfiable, there will be no $(\delta,\epsilon)$-far Nash equilibrium (both players have to play $f$ with a probability of at least $1-\epsilon$ and we know that $\delta>\epsilon$). 
\end{proof}

\begin{proposition}\label{deltahardfinal}
 Given any 3CNF formula $\phi$ with $n$ variables, for any $\delta$ polynomially bounded in $n$, there exists a game $\mathcal{D}(\phi,\delta,\epsilon)$ such that for some $\epsilon$ polynomially bounded in $n$, there exists a $(\delta,\epsilon)$-far Nash equilibrium in $\mathcal{D}(\phi,\delta,\epsilon)$ iff $\phi$ is satisfiable. 
\end{proposition}
\begin{proof}
    For $\delta>\frac{1}{n}-\frac{1}{n^2}-\frac{1}{n^3}$, the hardness result follows by  Proposition \ref{Hdelta} with the game $\mathcal{H}(\phi,\delta,\epsilon)$. Otherwise, we create a new game $\mathcal{D}(\phi,\delta,\epsilon)$ similar to the game $\mathcal{H}(\phi,\delta,\epsilon)$ with the difference that we make more copies of one literal.  For some literal $l^t\in L$ (and also its negation $-l^t$) from the original formula $\phi$, we make $d$ copies and $i$ out of these $d$ copies will be duplicated in a way that each player will be inclined to play only from the associated copies. This can be attained by using a  game structure similar to that discussed in  Proposition \ref{modifedconitzer} and the technique used in Lemma \ref{changelabellemma}.)

    We consider $L^{\mathcal{D}}=L \cup(\bigcup_{j=i}^{d} l^t_j)\cup(\bigcup_{j=i}^{d} (-l^t_j))-\{l^t,-l^t\}$ to be all literal strategies that are shared between the players where for each $j$, $l^t_j$ is just a copy of $l^{t}$. Similarly, $L^{\mathcal{D}}_1=(\bigcup_{j=1}^{i} l^t_{1,j})\cup(\bigcup_{j=1}^{i} -l^t_{1,j}) -\{l^t_1,-l^1_t\} $ and $L^{\mathcal{D}}_2=(\bigcup_{j=1}^{i} l^t_{2,j})\cup ( \bigcup_{j=1}^{i} -l^t_{2,j} )-\{l^t_2,-l^t_2\} $ will be considered as the sets that contain the literals that are associated with the row and the column player respectively. Finally, we define the strategy set of the game $\mathcal{D}(\phi,\delta,\epsilon)$ to be $S^\mathcal{D} = L^{\mathcal{D}}\cup L^{\mathcal{D}}_1 \cup L^{\mathcal{D}}_2 \cup V \cup C \cup\{f\}$.

    In this game, for the row player, the summation of the probabilities assigned to literals that are copies of  $l^t$ and $-l^t$ of the original formula, must be at least $\frac{1}{n}-2\epsilon$ followed by Lemma  \ref{sum1ndisjoint}, \ref{fmodifiedconitzer}, and  \ref{lornegativel} (if both players play $f$ with a positive probability less than $\frac{\epsilon}{n}$). In addition, the summation of the probabilities that are assigned to  literals that are copies of $l^t$ or $-l^t$ must be at least $\frac{1}{n}-\frac{1}{n^2}-2\epsilon$ while the other becomes less than $\frac{1}{n^2}$.  
   
    We set the payoff $u^{\mathcal{D}}_1(f,f)=u^{\mathcal{D}}_2(f,f)$ to be high enough so that all $\epsilon$-Nash equilibria that include the case that both players play $f$ cannot form a $\delta$-far Nash equilibrium (see \ref{uniqueanddisjoint}). In conclusion, if the formula $\phi$ is unsatisfiable, there will be no $(\delta,\epsilon)$-far Nash equilibrium. If the formula $\phi$ is satisfiable, then we can find a $\frac{i}{d}(\frac{1}{n}-\frac{1}{n^2}-\frac{1}{n^3})$-far Nash equilibrium (by spreading the weight evenly among the copies of $l^t$) and set $i$ to be small enough so that this fraction becomes greater than the given $\delta$. 
\end{proof}

This concludes the fact that the problem of whether there exists a $(\delta,\epsilon)$-far Nash equilibrium is NP-hard. Our results also hold for stronger cases as follows:

\begin{corollary}
    The problem of whether there exists a Nash equilibrium $(x^*,y^*)$ for a bi-matrix game $\mathcal{G}$ with the following property is NP-complete.
    \begin{itemize}
        \item $ 2\alpha  \leq ||x^*-y^*||_1\leq 2\beta$ for some $1\geq\beta>\alpha\geq 0$ polynomially bounded in the size of the strategies of $\mathcal{G}$.
    \end{itemize}
\end{corollary}

\subsection{Existence of Far Nash and Theorem \ref{sdfarmin}}

We prove that a far Nash equilibrium exists for a sufficiently large approximation error. This follows by the existence of a Nash equilibrium and simple techniques. We begin with a lemma and then proceed to the proof.

\begin{lemma} \label{equalgames}
	Let $\mathcal{G}$ be a game that has a symmetric Nash equilibrium  $(z^*,z^*)$ such that some pure strategy $t$ has a probability greater than $\delta$ in $z^*$.
 Then, there is a strategy $z^-$ such that $(z^-,z^*)$ and $(z^*,z^-)$ are $2\delta$-approximate Nash equilibria and $||z^*-z^-||_1\ge 2\delta$.
 \end{lemma}

\begin{proof}
	In this game, we have at least one pair of strategies $(z^*,z^*)$ that forms a Nash equilibrium. For one player, we give  $\delta$  share of $z^*_t$ (arbitrary $t$) evenly to all other strategies to generate $z^-$.	This modification on strategies has the property that $z^-$  and $z^*$ have $2\delta$-farness. Then the strategy profile $(z^-,z^*)$ has the property $||z^--z^*||\geq 2\delta$. This also holds for $(z^*,z^-)$ by a symmetric argument. Both $(z^-,z^*)$ and $(z^*,z^-)$ are $2\delta$-approximate Nash equilibrium by Lemma \ref{blowup}.

\end{proof}

 \subsubsection{Proof of Theorem \ref{sdfarmin} }

\begin{proof}

 Suppose  $(x^*,y^*)$ is a Nash equilibrium in $\mathcal{G}$. If $||x^*-y^*||_1>2\delta$, we are finished with the problem. Otherwise, $||x^*-y^*||_1\leq 2\delta$, we can change $x^*$ to $y^*$ (or the other way around) so that there exists $z^*$ such that $(z^*,z^*)$ will form a $(0,2\delta)$-far equilibrium by an argument similar to Lemma \ref{blowup}. For the next step, we simply can use Lemma \ref{equalgames} to generate $z^-$ and $||z^{-}-z^{*}||_1= 2\delta$ .

By Lemma \ref{blowup} and \ref{equalgames}, we can conclude that $(z^{-},z^*)$ is a $4\delta$-approximate Nash equilibrium ( $||x^*-z^-||\leq 4\delta$) and we know  $||z^{-}-z^{*}||_1= 2\delta$ which indicates that we can generate $(\delta,4\delta)$-far Nash equilibrium. It is obvious that there exists one strategy that is assigned with probability greater than $\frac{1}{n}$ in each distribution and we can modify this strategy accordingly.  Our proof reduces this problem to an instance of approximate Nash equilibrium which means that this problem is in PPAD.

\end{proof}

\begin{remark}
    
    To get a better approximation factor compared to $4\delta$,  we could look at the following optimization problem:
\[Min_{x\in \Delta_n}||x-x^*||_1\]
\[Subject~ to~ ||x-y^*||_1\geq 2\delta\]

\end{remark}

\subsection{Constrained Far and Disjoint Equilibrium}

An EXACT constrained far equilibrium does not necessarily exist in all games. Consider the well-known \emph{rock-paper-scissors} which is a zero-sum game (table \ref{tab:table5}). This game has a unique Nash equilibrium where both players uniformly select all three pure strategies and put them in their support. For each strategy, the game has one \emph{anti-strategy} where one of the players gets $1$ when the player plays the anti-strategy while the opponent gets $-1$. If both players play the same pure strategy they both get $0$ payoff. There are two possible cases for each player if we limit the players to respect the constraint set $\mathcal{R}^{n}_\mathrm{disjoint}$. The first case is that both players pick one pure strategy and the second case happens when one player plays one pure strategy while the other player plays two pure strategies. In either case, both players prefer to play the anti-strategies as this option is always available for at least one player. So, this game does not have an EXACT constrained disjoint equilibrium. But for example, if one player chooses `rock' and the other player players `paper' with probability $1-\epsilon$ and `scissors' with probability $\epsilon$. We can form an $\epsilon$-constrained disjoint equilibrium.  We are now ready to prove Proposition \ref{polyconst}.

% \usepackage{graphicx}
\begin{table}[]
\centering
\caption{Rock-Paper-Scissors, a zero-sum game}
\label{tab:table5}
\resizebox{220px}{!}{%
\begin{tabular}{llll}
                              &                             & Player 1                    &                             \\ \cline{2-4} 
\multicolumn{1}{l|}{}         & \multicolumn{1}{l|}{(0,0)}  & \multicolumn{1}{l|}{(-1,1)} & \multicolumn{1}{l|}{(1,-1)} \\ \cline{2-4} 
\multicolumn{1}{l|}{Player 2} & \multicolumn{1}{l|}{(1,-1)} & \multicolumn{1}{l|}{(0,0)}  & \multicolumn{1}{l|}{(-1,1)} \\ \cline{2-4} 
\multicolumn{1}{l|}{}         & \multicolumn{1}{l|}{(-1,1)} & \multicolumn{1}{l|}{(1,-1)} & \multicolumn{1}{l|}{(0,0)}  \\ \cline{2-4} 
\end{tabular}%
}
\end{table}

\subsubsection{Proof of Proposition \ref{polyconst} }

\begin{proof}

      Choose a pure strategy $l$ and let that be the column player's strategy. The row player will select a  pure strategy $t$ (other than $l$) that maximizes row player's payoff. Let $x^*$ be the row player's strategy. For all $i\in[n]-\{t,l\}$, put $x^-_i=\frac{\epsilon}{(n-2)}$ and $x^-_t=1-\epsilon$.  By lemma \ref{blowup}, the row player can lose at most $\epsilon$. Therefore, $x^-$ is a $\epsilon$-best response under the disjointness constraint for the row player. In this situation, the column player cannot obtain an improved payoff, unless the column player plays strategies that are in the support of $x^-$ which violates the constraints. 
\end{proof}
\begin{remark}
    This result shows that the disjointness of strategies as a common (coupled) constraint is flawed. This is because one selfish player can degrade the social welfare of all other players if all players comply with the common constraint. This indeed provides a justification for considering $(\theta,\epsilon)$-restricted disjoint equilibrium.
\end{remark}

\subsection{Hardness of Constrained Far Equilibrium and Theorems \ref{farppadhard} and \ref{farinppad}}

In the presence of constraints such as disjointness and farness, the existence of a constrained equilibrium cannot be guaranteed.  However, we show that under some simple assumptions, a solution can be guaranteed.  The following definition identifies a class of games that we will show are guaranteed to have a constrained far equilibrium, under some conditions. It also allows us to establish a relationship between constrained far equilibrium and  far Nash equilibrium (see Proposition~\ref{noescape}).

\begin{definition}\label{refdmg}
For a game $\mathcal{G}$, the \emph{diagonally modified} version of $\mathcal{G}$, denoted $\mathcal{D^\text{M}(G)}$,  is a game where the entries of the payoff matrices   $U_i^{\mathcal{D^\text{M}(G)}}$ are the same as $\mathcal{G}$ except the diagonal entries as follows: 
	\[ \forall t\in S:U^{\mathcal{D^\text{M}(G)}}_i(t,t)= -M\]
	
\end{definition}
\begin{remark}
	   It is clear to see that when $M$ is large enough, the chances of having disjoint equilibria increases. Note that if we find a Nash equilibrium in  $\mathcal{D^{\text{M}}(G)}$, it does not necessarily form a Nash equilibrium for the original game $\mathcal{G}$ since we removed all the information that the diagonals of the payoff matrices carry.
\end{remark}

\subsubsection{Proof of Theorem \ref{farppadhard}}

The following proposition concludes the poof of Theorem \ref{farppadhard} for $\delta=1$.

\begin{proposition}\label{zerodiag}
For any $\epsilon$ polynomially bounded in the size of any game $\mathcal{G}$, there exists an $\epsilon^\prime$  polynomially bounded in the size of $\mathcal{G}$ such that the problem of finding $\epsilon$-Nash equilibrium in the bi-matrix game $\mathcal{G}$ is poly-time reducible to the problem of finding $\epsilon^\prime$-constrained disjoint equilibrium in $\mathcal{D^{\text{M}}(G)}$.
	
	\end{proposition}
 \begin{proof}
     The proof follows by \ref{changelabellemma}, \ref{blowup}, and the fact that the problem of finding a Nash equilibrium is a PPAD-hard by \cite{Chen}. A complete proof is given in Appendix C.
 \end{proof}

 \begin{remark}
     The proof of the previous proposition also implies that this modification on the diagonals entries, does not change the fact that finding a Nash equilibrium is still PPAD-hard.
 \end{remark}

\subsubsection{Essential Elements for Theorem  \ref{farinppad}}

In order to prove Theorem \ref{farinppad}, we need to prove some properties of \emph{diagonally modified games}. We begin with the following definition and establish then establish a relationship between this definition and constrained far equilibrium.

\begin{definition}\label{semidisjointdef}
    Suppose that we are given a bi-matrix game and the goal is to find a Nash equilibrium  with the following property:
    \begin{itemize}
        \item For all $t\in[n]$, if both players play $t$ with a positive probability, then at least one of the players has to play $t$ with probability smaller than $\frac{1}{M}$. \end{itemize}
   This equilibrium is called \emph{$M$-semi-disjoint Nash} equilibrium. A \emph{constrained $M$-semi-disjoint} equilibrium can be defined in a similar manner where we only investigate deviations with respect to strategies that have the property.
\end{definition}

\begin{lemma} \label{guranteedsol1}
Suppose that any in Nash equilibrium of $\mathcal{D^{\text{M}}(G)}$, at least one of the players does not play a fully mixed strategy. For any pure strategy $t$ that is played with a positive probability by both two players, there exists at least one player that has to play strategy $t$ with probability less than $\frac{1}{M}$. Furthermore, for $\epsilon$-approximate Nash, this holds for probability less than $\frac{1+\epsilon}{M}$ .
\end{lemma}

\begin{proof}

By the assumption, for any Nash equilibrium of $\mathcal{D^{\text{M}}(G)}$, at least one player does not play a fully mixed strategy. Assume that for one Nash equilibrium $(x^*,y^*)$, there exists a pure strategy $t$ such that $y^*_t>\frac{1}{M}$  and $x^*_t>\frac{1}{M}$. Without loss of generality, assume that the column player does not play $t^\prime \in [n]$. We obtain a contradiction by showing that the row player can deviate from $t$ to $t^\prime$ to obtain a better payoff. Letting $(t,y^*)$ denote the strategy profile in which the row player plays the pure strategy $t$ with probability 1 and the column player plays the mixed strategy $y^*$, we have:
 \begin{equation}\label{paritalt}
     U^{\mathcal{D^{\text{M}}(G)}}_1(t,y^*)=\sum_{1 \leq j\leq n }  y_j^* U^{\mathcal{D^{\text{M}}(G)}}_1(t,j) < 1+ y_t^* U^{\mathcal{D^{\text{M}}(G)}}_1(t,t)   \leq 0 
 \end{equation}

where last inequality comes from the fact that $y^*_t\geq \frac{1}{M}$. Assuming $y^*$ is fixed, if we let $U^{\mathcal{D^{\text{M}}(G)}}_1(x^*_t,y^*)$ (where  $x^*_t$ does not denote a mixed strategy but rather the probability with which the pure strategy $t$ is played in $x^*$)  denote $x^*_t \cdot U^{\mathcal{D^{\text{M}}(G)}}_1(t,y^*)$, we have:
	\begin{equation}\label{paritalt2}
	    U^{\mathcal{D^{\text{M}}(G)}}_1(x^*_t,y^*)=x^*_t\sum_{1 \leq j\leq n }  y_j^* U^{\mathcal{D^{\text{M}}(G)}}_1(t,j)\leq x^*_t(1+y_t^* U^{\mathcal{D^{\text{M}}(G)}}_1(t,t))< 0
	\end{equation}

    Now suppose that the row player moves all the weight $x^*_t$ from $t$ to $t^\prime$. Call the modified strategy $x^{\prime}$. It is obvious that $U^{\mathcal{D^{\text{M}}(G)}}_1(x^{\prime}_{t^\prime},y^*)\geq 0$ because $y^*_{t^\prime}=0$.
	\[U^{\mathcal{D^{\text{M}}(G)}}_1(x^*,y^*)-U^{\mathcal{D^{\text{M}}(G)}}_1(x^{\prime},y^*)=x^*_t[\sum_{j=1}^{n}y^*_jU^{\mathcal{D^{\text{M}}(G)}}_1(t,j)-\sum_{j=1}^{n}y^*_jU^{\mathcal{D^{\text{M}}(G)}}_1(t^\prime,j)] \]
	\[U^{\mathcal{D^{\text{M}}(G)}}_1(x^*,y^*)-U^{\mathcal{D^{\text{M}}(G)}}_1(x^{\prime},y^*)= U^{\mathcal{D^{\text{M}}(G)}}_1(x^{*}_{t},y^*) - U^{\mathcal{D^{\text{M}}(G)}}_1(x^{\prime}_{t^\prime},y^*)< 0\]
		
In conclusion, assuming that for all $j$, we have $ y_j^*>\frac{1}{M}$, we reach a contradiction. We can easily extend the proof to work with the approximate Nash equilibrium as well.
\end{proof}

\begin{proposition}\label{inppad}
    For a diagonally modified game  $\mathcal{D^{\text{M}}(G)}$ where at least one player does not play a fully mixed strategy, the problem of finding a $M$-semi-disjoint Nash equilibrium is in PPAD. 
\end{proposition}
\begin{proof}
By Lemma~\ref{guranteedsol1}, \emph{any} Nash equilibrium of $\mathcal{D^{\text{M}}(G)}$  satisfying the condition is a $M$-semi-disjoint Nash equilibrium.
\end{proof}
\begin{proposition}\label{farandsemi}
   For any bi-matrix game $\mathcal{G}$, a $M$-semi constrained disjoint equilibrium is a $(1-\frac{n}{M})$ constrained far equilibrium.
\end{proposition}
\begin{proof}
    The proof  follows immediately by the definition of constrained far equilibrium.
\end{proof}

\subsubsection{Proof of Theorem \ref{farinppad}}
\begin{proof}
   Suppose $\mathcal{G}$ is a game whose diagonally modified version has no fully-mixed Nash equilibrium. We first show that the problem of finding a $M$-semi constrained disjoint equilibrium with approximation error $\frac{6n}{M}$ is in PPAD. Next, by Proposition \ref{farandsemi}, the problem of finding a constrained $(1-\frac{n}{M},\frac{6n}{M})$-far equilibrium is in PPAD.

Recall that the problem of finding an EXACT $M$-semi disjoint Nash is in PPAD for diagonally modified games if the diagonally modified game does not have a fully mixed Nash equilibrium. Let $\mathcal{S}^{n}_\mathrm{M}(y^*)$ be the row player's constraint set for a $M$-semi constrained disjoint equilibrium. Now suppose that $(x^*,y^*)$ is an EXACT $M$-semi disjoint Nash for $\mathcal{D^{\text{M}}(G)}$. To prove that $(x^*,y^*)$ is a $M$-semi constrained disjoint equilibrium with approximation error $\frac{6n}{M}$ in $\mathcal{G}$, we need to prove the following for all $f\in \mathcal{S}^{n}_\mathrm{M}(y^*)$ (The proof for the column player is similar):
\[  U^{\mathcal{G}}_1(x^*,y^*)+\frac{6n}{M}>U^{\mathcal{G}}_1(f,y^*) \]

Write  $x^*=x^b+x^s$ and $y^*=y^b+y^s$ where the superscript $b$ includes the strategies that are used with probability greater than $\frac{1}{M}$ in $x^*$ while the superscript $s$ is used for the other strategies (similarly for $y^b$ and $y^s$.) It is obvious that $U^{\mathcal{D^{\text{M}}(G)}}_1(x^b,y^b)=U^{\mathcal{G}}_1(x^b,y^b)$ since $x^b$ and $y^b$ both are associated with strategies that have disjoint supports. Since  $(x^*,y^*)$ is a $M$-semi  disjoint Nash equilibrium for $\mathcal{D^{\text{M}}(G)}$, it will satisfy the following:
 \begin{equation}\label{equaa}
     \forall f\in \mathcal{S}^{n}_\mathrm{M}(y^*) \subset \Delta_n, \quad
 U^{\mathcal{D^{\text{M}}(G)}}_1(x^*,y^*)\geq U^{\mathcal{D^{\text{M}}(G)}}_1(f,y^*)
 \end{equation}

 The following inequality holds by applying Lemma \ref{blowup} to Equation \ref{equaa}. If we write $\mathcal{D^{\text{M}}(G)}_1(x^b,y^*)$ to denote the share of the row player's payoff coming from the sub-distribution $x^b$, we have:
\[ \forall f\in \mathcal{S}^{n}_\mathrm{M}(y^*) \subset \Delta_n, \quad
 U^\mathcal{D^{\text{M}}(G)}_1(x^b,y^*)+\frac{n}{M}\geq U^{\mathcal{D^{\text{M}}(G)}}_1(f,y^*) \]
Similarly:
\[ \forall f\in \mathcal{S}^{n}_\mathrm{M}(y^*) \subset \Delta_n, \quad
 U^{\mathcal{G}}_1(x^b,y^b)+\frac{n}{M}+\frac{n}{M} \geq U^{\mathcal{D^{\text{M}}(G)}}_1(f,y^*) \]
and by the fact that $|U^{\mathcal{G}}_1(f,y^b)-U^{\mathcal{D^{\text{M}}(G)}}_1(f,y^*)|\leq \frac{n}{M}$:
\[ \forall f\in \mathcal{S}^{n}_\mathrm{M}(y^*) \subset \Delta_n, \quad
 U^{\mathcal{G}}_1(x^b,y^b)+\frac{2n}{M}+\frac{n}{M} \geq U^{\mathcal{G}}_1(f,y^b) \]
 
Now, we select one of the elements in the support of $x^b$ and $y^b$ and give them the summation of all of the probabilities in $x^s$ and $y^s$ respectively. These two new distributions $x^\prime$ and $y^\prime$  form a $M$-semi constrained disjoint equilibrium with $\frac{6n}{M}$ as the approximation error by \ref{blowup} (since $||x^\prime-x^b||_1\leq \frac{n}{M}$ and   $||y^\prime-y^b||_1\leq \frac{n}{M}$ ):
\[ \forall f\in \mathcal{S}^{n}_\mathrm{M}(y^\prime) \subset \Delta_n, \quad
 U^{\mathcal{G}}_1(x^\prime,y^\prime)+\frac{6n}{M} \geq U^{\mathcal{G}}_1(f,y^\prime) \]
\end{proof}

\subsection{An Interesting Connection}\label{dmg}
The following proposition shows how diagonally modified games provide a connection between disjoint equilibria in the constrained equilibrium and Nash equilibrium settings.
\begin{proposition}\label{noescape}
 For $\theta\geq \frac{1}{M}$, any $(\theta,\epsilon)$-restricted  disjoint equilibrium in $\mathcal{G}$ forms an  $\epsilon$-approximate disjoint Nash equilibrium in $\mathcal{D^{\text{M}}(G)}$. 
\end{proposition}
\begin{proof}
       See Appendix C.
\end{proof}

\subsection{Restricted Far Equilibrium and Proof of Theorem \ref{disjointrestricted} and \ref{farrestrict2}}

We begin with the following proposition that establishes a relationship between disjoint Nash equilibrium and constrained disjoint equilibrium.

\subsubsection{Essential Elements to Prove Theorem \ref{farrestrict2}}	

In order to prove Theorem \ref{farrestrict2}, we begin with a generalization of \ref{modifedconitzer} by embedding a zero-sum sub-game into the game. We will modify $\mathcal{G}(\phi,\epsilon)$ to $\mathcal{C}(\phi,\epsilon)$ ($\frac{1}{\epsilon}>c> n$) by removing rules 10-15 and adding a series of rules instead. Let $S^{\mathcal{C}} \equiv S^{\mathcal{C}}_1=S^{\mathcal{C}}_2=L_1 \cup L_2 \cup V \cup C \cup F^{\mathcal{C}} $ be the strategy set for both players where $F^{\mathcal{C}}=\{f^i~|~i\in[c]\}$. The new rules are defined as follows:	
	\begin{enumerate}
	
	    \item $u^{\mathcal{C}}_{1}(x, f^i)=u^{\mathcal{C}}_{2}(f^j, x)=0$ for all $i,j\in [c]$ and $x \in S^{\mathcal{C}} -F^{\mathcal{C}}$;
	      \item $u^{\mathcal{C}}_{1}(f^i, x)=u^{\mathcal{C}}_{2}(x, f^i)=n-1$ for all $i,j\in [c]$ and $x \in S^{\mathcal{C}}-F^{\mathcal{C}}$;
	    \item $u^{\mathcal{C}}_{1}(f^i, f^j)=u^{\mathcal{C}}_{2}(f^j, f^i)=\frac{n^2}{\epsilon}$ for all $i=j\in [c]$;
	      \item $u^{\mathcal{C}}_{1}(f^i, f^j)=u^{\mathcal{C}}_{2}(f^j, f^i)=2(\frac{n^2}{\epsilon})$ for all $i, j\in [c]$ and $i = j+1 \pmod{c}$;
	        \item $u^{\mathcal{C}}_{1}(f^i, f^j)=u^{\mathcal{C}}_{2}(f^j, f^i)=0$ for all $i, j\in [c]$ and $i \neq j+1 \pmod{c}$;

	 \item $
u^{\mathcal{C}}_{1}(l^2,*)=u^{\mathcal{C}}_{2}(*,l^1)=-2n \thickspace \mbox{for all} \thickspace *\in S^{\mathcal{C}},~  l^1 \in L_1 ~\text{and}~ l^2 \in L_2
$;
\item $
u^{\mathcal{C}}_{2}(l^2,*)=u^{\mathcal{C}}_{1}(*,l^1)=0 \thickspace \mbox{for all} \thickspace *\in S^{\mathcal{C}}-F^{\mathcal{C}},~  l^1 \in L_1 ~\text{and}~ l^2 \in L_2
$;
	\end{enumerate}

\begin{proposition}\label{disjointrestrict}
	Given an instance of \text{3CNF} $\phi$, $\mathcal{C}(\phi,\epsilon)$ has  an EXACT disjoint Nash equilibrium in which all strategies that are in the support of either of the players are played with probability strictly greater than $\frac{1}{c}$  iff the given formula $\phi$ is satisfiable. 
\end{proposition} 
	\begin{proof}
	When the formula is satisfiable a solution similar to \ref{modifedconitzer} can be attained where all players play some literals with probability of $\frac{1}{n}$. We need to show that when a formula is not satisfiable, the only EXACT Nash equilibrium of the game $\mathcal{C}(\phi,\epsilon)$ can have compared to $\mathcal{G}(\phi,\epsilon)$ is when players play the following strategy:
	 \[x^c=y^c=(\frac{1}{c}f_1,\ldots,\frac{1}{c}f_c)\]

It is easy to check that $(x^c,y^c)$ is an EXACT Nash equilibrium  since the sub-game induced by strategies in $F^{\mathcal{C}}$ is just a simple generalization of \emph{rock-paper-scissors}. It is also not hard to see that in any EXACT Nash equilibrium if none of the players play any strategy in $F^{\mathcal{C}}$, all strategies in $C\cup V$ cannot be played with a positive probability by an analysis similar to \cite{Conitzer} and Lemma \ref{maxsw2} since social welfare with the use of these strategies must be less than $2n-2$ and at least one of the players will try to achieve $n-1$ by deviating to any $f^i\in F^{\mathcal{C}}$. Using any strategy $f^i\in F^{\mathcal{C}}$ will cause the opponent player to play $f^{j}$ where $ j=i+1 \pmod{c}$ and this can continue forever.

Without loss of generality, suppose that the row player plays one strategy $f^i\in F^{\mathcal{C}}$ with probability greater than $\frac{\epsilon}{n}$.  A simple calculation shows that the expected payoff of the column player playing the pure strategy  $f^j\in F^{\mathcal{C}}$  where $ j=i+1\pmod{c}$ is always greater than $n$ ($f^i$ cannot be played with a minor probability). This indeed will prevent the column player from playing any strategy from $L$, $C$, or $V$. The sub-game $\mathcal{C}_F$ induced by considering the strategy set to be $F^{\mathcal{C}}\subset S^{\mathcal{C}}$   has only one EXACT unique Nash equilibrium $(x^c,y^c)$. For this strategy profile, all strategies in the support of both players are playing with a probability of at most $\frac{1}{c}$.
\end{proof}
\begin{remark}
\label{remarkofrestricted}
 We can also observe that the expected guaranteed payoff of any Nash equilibrium of $\mathcal{C}(\phi,\epsilon)$  is $\frac{1}{c}\cdot\frac{n^2}{\epsilon}+\frac{1}{c}\cdot\frac{2n^2}{\epsilon}=\frac{3n^2}{c\epsilon}$.   
\end{remark}

\subsubsection{Proof of Theorem \ref{disjointrestricted} and \ref{farrestrict2}}
The following proposition concludes the proof of Theorem $\ref{disjointrestricted}$.
\begin{proposition}\label{farrestrict}
Given a formula $\phi$ with $n$ variables, there exists a game such that for some $\theta$ and $\epsilon$  polynomially bounded in $n$, there exists a $(\theta,\epsilon)$-restricted disjoint equilibrium  in this game iff $\phi$ is satisfiable.
 \end{proposition}

\begin{proof}
It is not hard to extend Proposition \ref{disjointrestrict} to the approximate case (with $\epsilon=\frac{1}{2n^3}$) by combining it with the techniques used in the proof of Proposition \ref{modifedconitzer}. We duplicate clauses, variables, and the strategies in $F^{\mathcal{C}}$ similarly to what we did in game $\mathcal{G}(\phi,\epsilon)$ and consider four functions that map copies of the associated strategies.  We also need to assume that $\epsilon \leq \frac{1}{c}-\frac{2\epsilon^2}{n}\leq \frac{1}{n}-2\epsilon-\frac{1}{n^2}$. We can obviously see that $\frac{1}{c}-\frac{2\epsilon^2}{n}$ is greater than $\epsilon$, $\frac{\epsilon}{n}$ and $\frac{2\epsilon}{n}+\frac{\epsilon^2}{2n^2}$. 

Each player can play unassociated literals, clauses, and variables only with probability smaller than $\frac{\epsilon}{n}$ since a constrained disjoint equilibrium is also a disjoint Nash equilibrium in this game.  This indicates that these strategies cannot be played in any $(\frac{1}{n}-2\epsilon-\frac{1}{n^2},\epsilon)$-restricted disjoint equilibrium. This statement also holds for the (associated copies of) variables and clauses that can be played only with probability less than $\epsilon$ by Lemma \ref{maxsw2} and \ref{uniqueanddisjoint}. 
   
      If none of the players play strategies in $F^{\mathcal{C}}_1$ or $F^{\mathcal{C}}_2$, an analysis similar to Proposition \ref{modifedconitzer} can show that at least one of the players will be motivated to select some of the strategies in $F^{\mathcal{C}}_1$ or $F^{\mathcal{C}}_2$. If the players are allowed to play some strategies from $F^{\mathcal{C}}_1$ or $F^{\mathcal{C}}_2$, we will prove that the analysis of Proposition \ref{disjointrestrict}  with minor adjustments still holds.  In this modified version of $\mathcal{C}(\phi,\epsilon)$ (with duplicated variables and clauses), in any EXACT constrained disjoint equilibrium with the condition that at least one of the players plays strategies from $F^{\mathcal{C}}_1$ or $F^{\mathcal{C}}_2$ with probability greater than $\frac{\epsilon}{2n}$, no associated literal, clause, or variable strategy can be played with a positive probability. In the approximate case,  literals, clauses, or variables can be played with probability of at most $\frac{\epsilon^2}{n^2}$. This is because we know there exists one strategy from $F^{\mathcal{C}}_1$ or $F^{\mathcal{C}}_2$ with probability greater than $\frac{2\epsilon}{n}$ and playing $f^{i+1\pmod{c}}$ (from $F^{\mathcal{C}}_1$ or $F^{\mathcal{C}}_2$ depending on the player) will provide $\epsilon$ difference in the payoff at least.  In conclusion, these strategies cannot be played in any $(\frac{1}{n}-2\epsilon-\frac{1}{n^2},\epsilon)$-restricted disjoint equilibrium.

  Recall that each player must guarantee a payoff of $\frac{3n^2}{c\epsilon}$ by Remark~\ref{remarkofrestricted} (and $(3\frac{n^2}{c\epsilon})-\epsilon$ for the approximate version). Finally, if a player prefers to play any strategy from $F^{\mathcal{C}}_1$ or $F^{\mathcal{C}}_2$ with probability smaller than $\frac{1}{c}-\frac{2\epsilon^2}{n}$, then there exists at least one strategy in $F^{\mathcal{C}}_1$ or $F^{\mathcal{C}}_2$ such that this player plays the strategy with probability greater than $\frac{1}{c}+\frac{2\epsilon^2}{n^2}$ since we showed that the first and the second player can only use strategies in  $F^{\mathcal{C}}_1$ and $F^{\mathcal{C}}_2$ in the restricted equilibrium respectively. The opponent can get a payoff of at least $(\frac{3n^2}{c\epsilon})+\frac{2\epsilon^2}{n^2}(\frac{n^2}{\epsilon})$ which is equal to $\frac{3n^2}{c\epsilon}+2\epsilon$.  This follows by the fact that the structure of the sub-game that is induced by strategies in $F^{\mathcal{C}}_1$ and $F^{\mathcal{C}}_2$ is a game similar to \emph{rock-paper-scissors}.

  The maximum social welfare of any equilibrium of this sub-game is $2\frac{3n^2}{c\epsilon}$ and this shows that one player cannot meet the specified guarantee  $\frac{3n^2}{c\epsilon}-\epsilon$. In conclusion, in this case, there will be no $\epsilon$-constrained disjoint equilibrium with a minimum probability greater than $\frac{1}{c}-\frac{2\epsilon^2}{n}<\frac{1}{n}-\epsilon-\frac{1}{n^2}$. This completes the proof that an arbitrary formula $\phi$ is satisfiable iff this modified version of $\mathcal{C}(\phi,\epsilon)$ has a $(\frac{1}{n}-\epsilon-\frac{1}{n^2},\epsilon)$-restricted disjoint equilibrium.
      
\end{proof}

\begin{proposition}\label{restrictedfar}
Given a formula $\phi$ with $n$ variables, there exists a game $\mathcal{R}(\phi,\delta,\epsilon)$ such that for some $\epsilon$,$\theta$ and for any given $\delta$ polynomially bounded in $n$,  there exists a restricted $(\theta,\delta,\epsilon)$-far equilibrium for $\mathcal{R}(\phi,\delta,\epsilon)$ iff $\phi$ is satisfiable.  
\end{proposition}
%{\mathcal{D}}
\begin{proof}

We only prove the theorem for $\delta<\frac{1}{n}-\frac{1}{n^2}-\frac{1}{n^3}$ and the other case is similar (see Proposition \ref{Hdelta}). We generate the final game $\mathcal{R}(\phi,\delta,\epsilon)$ which is a combination of the games $\mathcal{D}(\phi,\delta,\epsilon)$ and $\mathcal{C}(\phi,\epsilon)$. This game has the strategy set $S^\mathcal{R}=L^{\mathcal{D}}\cup L^{\mathcal{D}}_1\cup L^{\mathcal{D}}_2\cup V_1 \cup V_2 \cup C_1 \cup C_2 \cup F^{\mathcal{R}}_1 \cup F^{\mathcal{R}}_2$. The literal strategies of this game are the same as the game $\mathcal{D}(\phi,\delta,\epsilon)$ and we duplicated the clauses and variables similar to $\mathcal{C}(\phi,\epsilon)$.  The function $\mathrm{g_1}:L^{\mathcal{D}}\cup L^{\mathcal{D}}_1\cup L^{\mathcal{D}}_2\rightarrow L$ is defined similarly based on $\mathrm{g}^\prime$ in the game $\mathcal{D}(\phi,\delta,\epsilon)$. We also have three functions $\mathrm{g_2}:V_1\cup V_2 \rightarrow V$ and $\mathrm{g_3}:C_1\cup C_2\rightarrow C$ (similar to $\mathrm{g_1}$, $\mathrm{g}$ and $\mathrm{g}^\prime$) that map the copied strategies to their original variables and clauses in $\phi$. The function $\mathrm{g_4}:F^{\mathcal{R}}_1\cup F^{\mathcal{R}}_2 \rightarrow F^{\mathcal{C}}$ can be used for strategies in $F^{\mathcal{C}}$.  The payoff matrices of the  game $\mathcal{R}(\phi,\delta,\epsilon)$ are defined as follows:	
	\begin{enumerate}
     \item $u^{\mathcal{R}}_{1}\left(l^1, l^2\right)=u^{\mathcal{R}}_{2}\left(l^2, l^1\right)=n-1$ for all $l^1,l^2 \in L^{\mathcal{D}}$ such that  $\mathrm{g_1}(l^1)\neq -\mathrm{g_1}(l^2)$ ;
 
	\item $u^{\mathcal{R}}_{1}\left(l^{1}, l^{2}\right)=u^{\mathcal{R}}_{2}\left(l^{1}, l^{2}\right)=n-1$ for all $l^{1} \in L^{\mathcal{D}}_1 ,~ l^{2} \in L^{\mathcal{D}}_2$ with $\mathrm{g_1}(l^{1} )\neq-\mathrm{g_1}(l^{2})$;

\item $u^{\mathcal{R}}_{1}\left(l^{1}, l^{2}\right)=u^{\mathcal{R}}_{2}\left(l^{1}, l^{2}\right)=n-1$ for all $l^{1} \in L^{\mathcal{D}},~ l^{2} \in L^{\mathcal{D}}_2$ with $\mathrm{g_1}(l^{1} )\neq-\mathrm{g_1}(l^{2})$;

\item $u^{\mathcal{R}}_{1}\left(l^{1}, l^{2}\right)=u^{\mathcal{R}}_{2}\left(l^{1}, l^{2}\right)=n-1$ for all $l^{1} \in L^{\mathcal{D}}_1,~ l^{2} \in L^{\mathcal{D}}$ with $\mathrm{g_1}(l^{1} )\neq-\mathrm{g_1}(l^{2})$;
  \item $u^{\mathcal{R}}_{1}(-l, l)=u^{\mathcal{R}}_{2}(-l, l)=n-4$ for all $l \in L^{\mathcal{D}}$;
	
  \item $u^{\mathcal{R}}_{1}(-l^1, l^2)=u^{\mathcal{R}}_{2}(-l^1, l^2)=n-4$ for all $l^{1}\in L^{\mathcal{D}}_1 ~, l^{2} \in L^{\mathcal{D}}_2$ and $\mathrm{g_1}(l^{1} )=\mathrm{g_1}(l^{2})$;
  \item $u^{\mathcal{R}}_{1}(-l^1, l^2)=u^{\mathcal{R}}_{2}(-l^1, l^2)=n-4$ for all $l^{1}\in L^{\mathcal{D}} ~, l^{2} \in L^{\mathcal{D}}_2$ and $\mathrm{g_1}(l^{1} )=\mathrm{g_1}(l^{2})$;
  
  \item $u^{\mathcal{R}}_{1}(-l^1, l^2)=u^{\mathcal{R}}_{2}(-l^1, l^2)=n-4$ for all $l^{1}\in L^{\mathcal{D}}_1 ~, l^{2} \in L^{\mathcal{D}}$ and $\mathrm{g_1}(l^{1} )=\mathrm{g_1}(l^{2})$;
	
	\item $u^{\mathcal{R}}_{1}(v^1, l)=u^{\mathcal{R}}_{2}(l, v^2)=n$ for all $v^1 \in V,~ v^2 \in V_2, ~ l \in L^{\mathcal{D}} $ with $\mathrm{g_2}(v^1)=g(v^2)$ and $\mathrm{v}(\mathrm{g_1}(l)) \neq \mathrm{g_2}(v^1) $;

 \item $u^{\mathcal{R}}_{1}(v^1, l^2)=u^{\mathcal{R}}_{2}(l^1, v^2)=n$ for all $v^1 \in V_1,~ v^2 \in V_2,~ l^1 \in L^{\mathcal{D}}_1, ~ l^2 \in L^{\mathcal{D}}_2$ with $ \mathrm{g_1}(l^1)=\mathrm{g_1}(l^2)$, $\mathrm{g_2}(v^1)=\mathrm{g_2}(v^2)$ and $\mathrm{v}(\mathrm{g_1}(l^1)) \neq \mathrm{g_2}(v^1)$;

 \item $u^{\mathcal{R}}_{1}(l, x^2)=u^{\mathcal{R}}_{2}(x^1, l)=n-4$ for all $l\in L^{\mathcal{D}}, ~ x^1 \in V_1\cup C_1, ~ x^2 \in V_2\cup C_2$ ;
 
	\item $u^{\mathcal{R}}_{1}(l^1, x^2)=u^{\mathcal{R}}_{2}(x^1, l^2)=n-4$ for all $l^{1} \in L^{\mathcal{D}}_1 ,~ l^{2} \in L^{\mathcal{D}}_2,~x^1 \in V_1\cup C_1, ~ x^2 \in V_2\cup C_2$ and $\mathrm{g_1}(l^1)=\mathrm{g_1}(l^2)$;

\item $u^{\mathcal{R}}_{1}(v^1, l)=u^{\mathcal{R}}_{2}(l, v^2)=0$ for all $v^1 \in V_1,~ v^2 \in V_2,~l\in L^{\mathcal{D}} $ with $\mathrm{g_2}(v^1)=\mathrm{g_2}(v^2)$ and $\mathrm{v}(\mathrm{g_1}(l))=\mathrm{g_2}(v^1)$;
	
	\item $u^{\mathcal{R}}_{1}(v^1, l^2)=u^{\mathcal{R}}_{2}(l^1, v^2)=0$ for all $v^1 \in V_1,~v^2\in V_2,~l^1 \in L^{\mathcal{D}}_1, ~ l^2 \in L^{\mathcal{D}}_2 $ with $\mathrm{g_1}(l^1)=\mathrm{g_1}(l^2)$, $\mathrm{g_2}(v^1)=\mathrm{g_2}(v^2)$ and $\mathrm{v}(\mathrm{g}(l^1))=\mathrm{g_2}(v^1)$;
 %x^1 \in S^{\mathcal{R}}-\{L^{\mathcal{D}},~ L^{\mathcal{D}}_1,~L^{\mathcal{D}}_2,V_2,C_2,F\}, ~ x^2 \in S^{\mathcal{R}}-\{L^{\mathcal{D}},~ L^{\mathcal{D}}_1,~L^{\mathcal{D}}_2,V_1,C_1,F\}
	
	\item $u^{\mathcal{R}}_{1}(v^1, x^2)=u^{\mathcal{R}}_{2}(x^1, v^2)=n-4$ for all $v^1 \in V_1,~ v^2\in V_2,~ x^1 \in V_1\cup C_1,~x^2 \in V_2\cup C_2$ and $\mathrm{g_2}(v^1)=\mathrm{g_2}(v^2)$;

    \item $u^{\mathcal{R}}_{1}(c^1, l)=u^{\mathcal{R}}_{2}(l, c^2)=n$ for all $c^1 \in C_1,~c^2 \in C_2,~ l\in L^{\mathcal{D}} $ with $\mathrm{g_3}(c^1)=\mathrm{g_3}(c^2)$ and $\mathrm{g_1}(l) \notin \mathrm{g_3}(c^1)$;
	
	\item $u^{\mathcal{R}}_{1}(c^1, l^2)=u^{\mathcal{R}}_{2}(l^1, c^2)=n$ for all $c^1 \in C_1,~c^2 \in C_2,~ l^1 \in L^{\mathcal{D}}_1, ~  l^2 \in L^{\mathcal{D}}_2$ with $\mathrm{g_1}(l^1)=\mathrm{g_1}(l^2)$, $\mathrm{g_3}(c^1)=\mathrm{g_3}(c^2)$ and $ \mathrm{g_1}(l^1) \notin \mathrm{g_3}(c^1)$;
	
	\item $u^{\mathcal{R}}_{1}(c^1, l)=u^{\mathcal{R}}_{2}(l, c^2)=0$ for all $c^1 \in C_1,~ c^2 \in C_2,~ l \in L^{\mathcal{D}}, ~ $  with $\mathrm{g_1}(l^1)=\mathrm{g_1}(l^2)$, $\mathrm{g_3}(c^1)=\mathrm{g_3}(c^2)$ and $ \mathrm{g_1}(l) \in \mathrm{g_3}(c^1)$;
 
 \item $u^{\mathcal{R}}_{1}(c^1, l^2)=u^{\mathcal{R}}_{2}(l^1, c^2)=0$ for all $c^1 \in C_1,~ c^2 \in C_2,~ l^1 \in L^{\mathcal{D}}_1, ~ l^2 \in L^{\mathcal{D}}_2$  with $\mathrm{g_1}(l^1)=\mathrm{g_1}(l^2)$ and $ \mathrm{g_1}(l^1) \in \mathrm{g_3}(c^1)$;

\item $u^{\mathcal{R}}_{1}(c^1, x^2)=u^{\mathcal{R}}_{2}(x^1, c^2)=n-4$ for all $v^1 \in V_1,~ v^2\in V_2,~ x^1 \in V_1\cup C_1,~x^2 \in V_2\cup C_2$ and $\mathrm{g_3}(c^1)=\mathrm{g_3}(c^2)$;

	    \item $u^{\mathcal{R}}_{1}(x, f^i)=u^{\mathcal{R}}_{2}(f^j, x)=0$ for any $f^i\in F^{\mathcal{R}}_2$ and $f^j\in F^{\mathcal{R}}_1$ where $x \in S^{\mathcal{R}} -F^{\mathcal{R}}_1-F^{\mathcal{R}}_2$;
	      \item $u^{\mathcal{R}}_{1}(f^i, x)=u^{\mathcal{R}}_{2}(x, f^j)=n-1$, for any $f^i\in F^{\mathcal{R}}_1$ and $f^j\in F^{\mathcal{R}}_2$  where  $x \in S^{\mathcal{R}}-F^{\mathcal{R}}_1-F^{\mathcal{R}}_2$;
	    \item $u^{\mathcal{R}}_{1}(f^i, f^j)=u^{\mathcal{R}}_{2}(f^j, f^i)=\frac{d n^2}{\delta\epsilon}$ for any $f^i\in F^{\mathcal{R}}_1$ and $f^j\in F^{\mathcal{R}}_2$ such that $\mathrm{g_4}(f^i)=\mathrm{g_4}(f^j)$ ;
     
	      \item $u^{\mathcal{R}}_{1}(f^i_1, f^j_2)=u^{\mathcal{R}}_{2}(f^j_1, f^i_2)=2(\frac{d n^2}{\delta\epsilon})$ for $f^i_1, f^j_1 \in F^{\mathcal{R}}_1$ and $f^i_2, f^j_2 \in F^{\mathcal{R}}_2$ where $\mathrm{g_4}(f^i_1)=\mathrm{g_4}(f^i_2)=f^i$ and $\mathrm{g_4}(f^j_1)=\mathrm{g_4}(f^j_2)=f^j$ (the superscripts $i$ and $j$ denote the $i$-th and $j$-th member of $F^\mathcal{C}$) while $i = j+1 \pmod{c}$;
   \item $u^{\mathcal{R}}_{1}(f^i_1, f^j_2)=u^{\mathcal{R}}_{2}(f^j_1, f^i_2)=0$ for $f^i_1, f^j_1 \in F^{\mathcal{R}}_1$ and $f^i_2, f^j_2 \in F^{\mathcal{R}}_2$ where $\mathrm{g_4}(f^i_1)=\mathrm{g_4}(f^i_2)=f^i$ and $\mathrm{g_4}(f^j_1)=\mathrm{g_4}(f^j_2)=f^j$ while $i \neq j+1 \pmod{c}$;

	 \item $
u^{\mathcal{R}}_{1}(l^2,*)=u^{\mathcal{R}}_{2}(*,l^1)=-2n \thickspace \mbox{for all} \thickspace *\in S^{\mathcal{R}},~  l^1 \in L_1 ~\text{and}~ l^2 \in L_2
$;
\item $
u^{\mathcal{R}}_{2}(l^2,*)=u^{\mathcal{R}}_{1}(*,l^1)=0 \thickspace \mbox{for all} \thickspace *\in S^{\mathcal{R}}-F^{\mathcal{R}}_1-F^{\mathcal{R}}_1,~  l^1 \in L_1 ~\text{and}~ l^2 \in L_2
$;

	 \item $
u^{\mathcal{R}}_{1}(v^2,*)=u^{\mathcal{R}}_{2}(*,v^1)=-2n \thickspace \mbox{for all} \thickspace *\in S^{\mathcal{R}},~  v^1 \in V_1 ~\text{and}~ v^2 \in V_2
$;
\item $
u^{\mathcal{R}}_{2}(v^2,*)=u^{\mathcal{R}}_{1}(*,v^1)=0 \thickspace \mbox{for all} \thickspace *\in S^{\mathcal{R}}-F^{\mathcal{R}}_1-F^{\mathcal{R}}_2,~  v^1 \in V_1 ~\text{and}~ v^2 \in V_2
$;

	 \item $
u^{\mathcal{R}}_{1}(c^2,*)=u^{\mathcal{R}}_{2}(*,c^1)=-2n \thickspace \mbox{for all} \thickspace *\in S^{\mathcal{R}},~  c^1 \in C_1 ~\text{and}~ c^2 \in C_2
$;
\item $
u^{\mathcal{R}}_{2}(c^2,*)=u^{\mathcal{R}}_{1}(*,c^1)=0 \thickspace \mbox{for all} \thickspace *\in S^{\mathcal{R}}-F^{\mathcal{R}}_1-F^{\mathcal{R}}_2,~  c^1 \in C_1 ~\text{and}~ c^2 \in C_2
$;

	 \item $
u^{\mathcal{R}}_{1}(f^2,*)=u^{\mathcal{R}}_{2}(*,f^1)=-2n \thickspace \mbox{for all} \thickspace *\in S^{\mathcal{R}},~  f^1 \in F^\mathcal{R}_1 ~\text{and}~ f^2\in F^\mathcal{R}_2
$;
\item $
u^{\mathcal{R}}_{2}(f^2,*)=u^{\mathcal{R}}_{1}(*,f^1)=0 \thickspace \mbox{for all} \thickspace *\in S^{\mathcal{R}}-F^\mathcal{R}_1 -F^\mathcal{R}_2 ,~  f^1 \in F^\mathcal{R}_1 ~\text{and}~ f^2\in F^\mathcal{R}_2.
$

	\end{enumerate}

%$\frac{i}{d}(\frac{1}{n}-\frac{1}{n^2}-\frac{1}{n^3})
 If the formula $\phi$ is satisfiable, then there exists a $(\frac{\delta}{d},\delta,0)$-restricted far equilibrium where $d$ is driven by $\delta$ (see Proposition \ref{deltahardfinal}). We show that if $\phi$ is not satisfiable, there exists no  $(\frac{\delta}{d}-\frac{2\delta\epsilon^2}{d n},\delta,\epsilon)$-restricted far equilibrium. If no (associated) strategy from $F^{\mathcal{R}}_1$ or $F^{\mathcal{R}}_2$ is played, one of the players cannot get the guaranteed payoff $n-1-\epsilon$ and will choose one strategy from these sets similar to all previous results.  If there exists one (associated) strategy in $F^{\mathcal{R}}_1$ or $F^{\mathcal{R}}_2$ such that the row or the column player play this strategy with probability greater than $\frac{\delta\epsilon}{2dn}$, all strategies that are related to literals, variables, and clauses cannot be played with probability greater than $\frac{\delta\epsilon^2}{dn^2}$ following  an analysis similar to Theorem \ref{farrestrict}.

  Finally, if a player prefers to play any strategy from $F^{\mathcal{R}}_1$ or $F^{\mathcal{R}}_2$ with probability smaller than $\frac{1}{c}-\frac{2\delta\epsilon^2}{dn}$, then there exists at least one strategy such that the opponent can get a payoff of at least $(\frac{3dn^2}{c\delta\epsilon})+\frac{2\delta\epsilon^2}{dn^2}(\frac{d n^2}{\delta\epsilon})$ which is equal to $\frac{3dn^2}{c\delta\epsilon}+2\epsilon$.  The maximum social welfare of any equilibrium of this game is sub-game $2\frac{3dn^2}{c\delta\epsilon}$ and this shows that one player cannot meet the guarantee  $\frac{3dn^2}{c\delta\epsilon}-\epsilon$. If we set $c=\lceil\frac{d}{\delta}\rceil+1$, then $\frac{1}{c}<\frac{\delta}{d}$ and $\frac{1}{c}-\frac{2\delta\epsilon^2}{dn^2}$ is always positive (generating $\mathcal{R}(\phi,\delta,\epsilon)$ with $|F^\mathcal{R}_1|=|F^\mathcal{R}_2|>\frac{\delta}{d}$). This completes the proof that an arbitrary formula $\phi$ is satisfiable iff the $\mathcal{R}(\phi,\delta,\epsilon)$ has a $(\frac{\delta}{d},\delta,\epsilon)$-restricted far equilibrium.

\end{proof}

\begin{corollary}\label{majornashhard}
    For a given formula $\phi$ with $n$ variables, there exists a game such that for some $\epsilon$ and $\theta$ polynomially bounded in $n$, a $(\theta,\epsilon)$-major Nash exists for this game iff the formula $\phi$ is satisfiable.
\end{corollary}

\section{Conclusion and Open Problems}
We have presented hardness and feasibility results for games with strategic constraints, as well as arguing that they are useful in modeling a variety of applications involving strategic behavior. We believe this is a fruitful area for further study with respect to computational issues as well as applications. Below we suggest some possible research directions.

In contrast to constrained disjoint equilibrium, even the inclusion of $(\delta,\epsilon)$-constrained far equilibrium in NP cannot be proved easily for any $\delta$. Even further, we showed that an equilibrium in this sense may not exist for zero-sum games. However, a \emph{constrained close equilibrium} with the following (convex) common constraint, has a guaranteed solution by \cite{rosen} which relies on Kakutani's Fixed Point Theorem: 
\[\mathcal{R}^{n}_\mathrm{\delta}=\{(x,y)\in \Delta_n\times\Delta_n~|~ ||x-y||_1\leq 2\delta \}\]
Inclusion in PPAD could be proven by using arguments similar to those discussed in \cite{concavegames}. Finding communication complexity and inapproximability results for constrained close equilibrium may also be of interest. In general, an interesting question is whether there exists a natural variant of  Nash equilibrium such that the computational complexity of the problem is in TFNP but not in $PPAD$. Finding a direct reduction for the hardness of EXACT partition Nash for bi-matrix games remains an open problem that does not seem to have a straightforward solution.

\bibliography{refs}

\appendix

\section{Technical Lemmas}

\subsection{Proof of Lemma \ref{blowup}}

\begin{proof}

	It is not hard to see $||x^--x^*||_1\leq 2\epsilon$.  We prove that we have $|U_1(x^*,y^*)-U_1(x^-,y^*)|\leq \epsilon(\beta-\alpha)$. Taking and redistributing (regardless of how we do so) $\epsilon_i$ from $x^*_i$ can lower a payoff at most $\epsilon_i(\beta-\alpha)$. So, the overall utility loss will be less than $\epsilon(\beta-\alpha)$.

	We show that $x^-$ is a $\epsilon(\beta-\alpha)$-best response to $y^*$ since  $x^*$ is a best response to $y^*$ which is proven by the following argument:
	\[U_1(x^*,y^*)\geq U_1(x,y^*)\]	\[U_1(x^-,y^*)+\epsilon(\beta-\alpha)\geq U_1(x,y^*)\]
	
We show  that $y^*$ is a $2\epsilon(\beta-\alpha)$-best response to $x^-$ . For all $y \in S$ we have:
	\[U_2(x^*,y^*)\geq U_2(x^*,y)\]
	\[U_2(x^-,y^*)+\epsilon(\beta-\alpha)\geq U_2(x^*,y)\geq  U_2(x^-,y) -\epsilon(\beta-\alpha) \thickspace (\text{for any pure strategy $y$})\]
		\[U_2(x^-,y^*)+2\epsilon(\beta-\alpha)\geq  U_2(x^-,y) \]
	This means $(x^-,y^*)$ is a $2\epsilon(\beta-\alpha)$-Nash equilibrium. We can prove this for $(x^*,y^-)$ similarly.

 Inequality (2) is established as follows:
	\[|U_1(x^*,y^*)-U_1(x^-,y^-)| \leq |U_1(x^*,y^*)-U_1(x^*,y^-)|+|U_1(x^*,y^-)-U_1(x^-,y^-)|
	\]
	\[\leq \epsilon (\beta-\alpha) +\epsilon (\beta-\alpha)=2\epsilon (\beta-\alpha)\]

 (3) is established similarly.
\end{proof}

\begin{remark}

	Let $\beta=1$ and $\alpha=0$. Here is an example that we use throughout the paper. We take $\delta$ fraction of pure strategy spread it equally to all other pure strategies.
	
	\begin{equation} \label{diffx}
		|U_1(x^*,y^*)-U_1(x^-,y^*)| = \left| \sum_{ 1 \leq  j \leq n}  \delta y^*_j U_1(t,j)- \sum_{ 1 \leq  i\neq t \leq n} \sum_{ 1 \leq  j \leq n} \frac{\delta}{(n-1)} y^*_j  U_1(i,j) \right|
	\end{equation}
	\[ =\left| \sum_{ 1 \leq  i\neq t \leq n}  \sum_{ 1 \leq  j \leq n}  \frac{\delta}{(n-1)} y^*_j U_1(t,j)- \sum_{ 1 \leq  i\neq t \leq n} \sum_{ 1 \leq  j \leq n} \frac{\delta}{(n-1)} y^*_j  U_1(i,j) \right|\]
	\[=\left| \frac{\delta}{(n-1)} \left(\sum_{ 1 \leq  i\neq t \leq n}  \sum_{ 1 \leq  j \leq n}   y^*_j U_1(t,j)- \sum_{ 1 \leq  i\neq t \leq n} \sum_{ 1 \leq  j \leq n}  y^*_j  U_1(i,j)\right) \right|\]
	
	It is obvious that $|U_1(t,j)-U_1(i,j)|\leq  1$ since we can assume games' payoff is restricted to $[0,1]$. Therefore:
	\[|U_1(x^*,y^*)-U_1(x^-,y^*)|\leq \left| \frac{\delta}{(n-1)} \sum_{ 1 \leq  i\neq t \leq n}  \sum_{ 1 \leq  j \leq n}   y^*_j   \right|\]

	Finally, since $y^*$ is a distribution::
	\[|U_1(x^*,y^*)-U_1(x^-,y^*)|\leq  \delta  \]
	We can do the same for the column player too:	\[	 |U_2(x^*,y^*)-U_2(x^*,y^-)|\leq \delta \]
	
	We calculated the exact difference between the utilities but the reason that why this forms a $2\epsilon$-approximate Nash equilibrium follows by \ref{blowup}. The calculations was done to show that even regulated modifications may cause a large approximation error.
\end{remark}

\subsection{Using  Support Enumeration  }\label{supportenusection}

\begin{remark}\label{matrixformapprx}
We can also define a Nash equilibrium for a bi-matrix game $\mathcal{G}(R,C)$ ($R$ and $C$ denote the payoff matrices for the row(first) and the column(second) player respectively) by doing the following calculations on matrices:
	\[
	\begin{aligned}
		x^{* T} R y^* & \geq e_{i}^{T} R y^*, \forall i \in\{1, \dots, m\} \\
		x^{* T} C y^* & \geq x^{* T} C e_{i}, \forall i \in\{1, \dots, n\}
	\end{aligned}
	\]
	
	We can also define an $\epsilon$-approximate Nash equilibrium by doing the following modification:
		\[
	\begin{aligned}
		x^{* T} R y^* +\epsilon & \geq e_{i}^{T} R y^*, \forall i \in\{1, \dots, m\} \\
		x^{* T} C y^* +\epsilon & \geq x^{* T} C e_{i}, \forall i \in\{1, \dots, n\}
	\end{aligned}
\]

\end{remark}\label{Matrixform}

\begin{proposition}
The mixed strategy $(x^*, y^*)$ is a Nash equilibrium of $\mathcal{G}(R, C)$ iff these two conditions are satisfied:
\[
\forall i \in Supp(x^*), (R y)_i=u=\max _{q \in S}\left\{(R y)_q\right\}
\]
\[\forall  j \in Supp(y^*), \left(x^T C\right)_j=v=\max _{r \in S}\left\{\left(x^T C\right)_r\right\}
\]

\end{proposition}

The first condition of this proposition states that a mixed strategy $x^*$ of the row player is a best response to mixed strategy $y^*$ of the column player if and only if all pure strategies $i$ in the support of $x^*$ are best responses to mixed strategy $y^*$. The second condition represents the best response condition corresponding to the column player.
\begin{proposition}\label{SUPPen}

The problem of whether a game has a Nash equilibrium in which both players play a fully mixed strategy is in P.
\end{proposition}

\begin{proof}
       The following linear equations can check the case whether both players can have a Nash equilibrium with support size $n$.  A candidate mixed strategy pair is determined by solving the equations:
\[
\begin{gathered}
\forall  j\in [n], ~ \sum_{i \in [n]} x_{i} c_{ij}=v,   \\
\sum_{i \in [n]} x_{i}=1
\end{gathered}
\]
and
\[
\begin{gathered}
\forall  j\in [n], ~  \sum_{j \in [n]} y_{j} r_{ij}=u   \\
\sum_{j \in [n]} y_{j}=1
\end{gathered}
\]
For the row player, the first set of equations tries to find a strategy  $x$ with support size $n$ that makes the column player indifferent among playing the pure strategies in $[n]$. That means the column player obtains the same payoff, $v$, by playing the pure strategies in $[n]$ if the row player  plays $x$. The second set of equations plays the same role for the column player. If the linear equations do not have a solution then there is no Nash equilibrium where both players play a fully mixed strategy. The systems of equations can be solved using back-substitution.
\end{proof}

\section{Complexity of Nash Equilibrium with a Guaranteed Payoff and Partition Nash}

Here we revisit a hardness result in \cite{vadhan} and break the proof down into multiple parts in a way that clarifies the connection to our techniques. The result shows that deciding whether a bi-matrix game has a $\epsilon$-Nash equilibrium with a guaranteed payoff of $n-1-\epsilon$ is NP-complete. We call this problem \emph{guaranteed Nash}.

\subsection{Approximate Guaranteed Nash}

The following definition of the game $\mathcal{SV}(\phi,\epsilon)$ from \cite{vadhan} (Theorem $8.6$) is simpler than our game $\mathcal{G(\phi,\epsilon)}$, as both players use the same literal set and $f$ was omitted from their strategy set in contrast to the game provided in \cite{Conitzer}.  The symmetric game $\mathcal{SV}(\phi,\epsilon)$ is defined as follows:

	\begin{enumerate}
	    \item $u_1\left(l^1, l^2\right)=n-1$, where $l^1 \neq -l^2$ for all $l^1, l^2 \in L$. This will ensure each player gets a high payoff for playing the aforementioned strategy.
	    \item $u_1(l, -l)=n-4$ for all $l \in L$. This will ensure that each player does not play a literal and its negation at the same time.
	    \item $u_1(v, l)=0$, where $\mathrm{v}(l)=v$, for all $v \in V, l \in L$. This, along with rule 4, ensures that for each variable $v$, each agent plays either $l$ or $-l$ with a probability of at least $\frac{1}{n}$, where $\mathrm{v}(l)=\mathrm{v}(-l)=v$.
	    
	    \item $u_1(v, l)=n$, where $\mathrm{v}(l) \neq v$, for all $v \in V, l \in L$.
	    
	    \item $u_1(l, x)=n-4$, where $l \in L,~ x \in V \cup C$. This, along with rules 6 and 7, ensures that if both players do not play the literals, then the payoffs cannot meet the guarantees.
	    \item $u_1(v, x)=n-4$ for all $v \in V, x \in V \cup C$.
	    \item $u_1(c, x)=n-4$ for all $c \in C, x \in V \cup C$.
	    \item $u_1(c, l)=0$ where $l \in c$ for all $c \in C, l \in L$. This, along with rule 9, ensures that for each clause $c$, each agent plays a literal in the clause $c$ with probability least $\frac{1}{n}$.
	    \item $u_1(c, l)=n$, where $l \notin c$ for all $c \in C, l \in L$.

	\end{enumerate}

 \begin{remark}
     The game introduced in both \cite{Conitzer,vadhan} is a symmetric game which means $\forall s_1,s_2 \in S,\space  u_1\left(s_1, s_2\right)=u_2\left(s_2, s_1\right) $.
 \end{remark}
	
\begin{proposition}\label{SVgame}
Let $\epsilon=\frac{1}{2 n^3}$ and let the guarantee to each player be $n-1$. Given a 3CNF $\phi$ is satisfiable iff there exists a $\epsilon$-Nash equilibrium in $\mathcal{SV}(\phi,\epsilon)$ where each player has a guaranteed payoff of $n-1-\epsilon$. 
\end{proposition}
\begin{proof}
We will prove this theorem in section \ref{SVproof}.
\end{proof}

\subsubsection{Essential Lemmas for Proposition \ref{SVgame}}
\begin{lemma}
	\label{maxsw}
    In any $\epsilon$-Nash equilibrium with the guaranteed payoff $n-1-\epsilon$ in $\mathcal{SV}(\phi,\epsilon)$, clauses and variables are played with a probability of at most $\epsilon$.
\end{lemma}
\begin{proof}
		The social welfare of $\mathcal{SV}(\phi)$ is at most $2n-2$. If neither player plays from the $L$, the social welfare is at most $2n-4$. When both players play from $L$  with probability $1-\epsilon$, the expected social welfare is at most $2n-2-2\epsilon$. So, for any $\epsilon^{\prime} > \epsilon$, playing variables and clauses with  probability $\epsilon^{\prime}$ will give an expected social welfare of at most $2n-2-2\epsilon'<2n-2-2\epsilon$. This means at least one player will have an expected payoff less than $n-1-\epsilon$ which violates the definition $\epsilon$-best response.	
	
\end{proof}
\begin{lemma}\label{lornegativel}
     In any $\epsilon$-Nash equilibrium with a guaranteed payoff of $n-1-\epsilon$ in $\mathcal{SV}(\phi,\epsilon)$, for any $l\in L$, the probability that the row player plays $l$ or $-l$ is at least  $\frac{1}{n}-2 \epsilon$.
\end{lemma}

\begin{proof}
    	Suppose this is not correct and there exists one $l \in L$ and the probability of $l$ being played will be less than $ \frac{1}{n} -\epsilon-\frac{2 \epsilon}{n} \geq \frac{1}{n}-2 \epsilon $. Then, the expected payoff for the column player playing the pure strategy $\mathrm{v}(l)$ is at least the summation of these cases:
	
	\begin{itemize}
		\item 	When the row player plays $l$ or $-l$ : $\left(\frac{1}{n}-\epsilon-\frac{2 \epsilon}{n}\right) \cdot 0$.
		
		\item When the row player plays a literal other than  $l$ or $-l$:
		\[(1-\epsilon-(\frac{1}{n}-\epsilon-\frac{2 \epsilon}{n}))\cdot n\]
		
		\item When the row player does not play a literal from $L$ the expected payoff portion of this possibility will be at least: $\epsilon \cdot 0$
	\end{itemize}
	The summation will be at least $n-1+2 \epsilon$. Since the maximum social welfare is $2n-2$, the other (row) player fails to meet the guarantee $n-1-\epsilon$. This is correct for the other player as well.
\end{proof}

\begin{lemma}\label{bothliterals}
      In any $\epsilon$-Nash equilibrium of $\mathcal{SV}(\phi,\epsilon)$ with a guaranteed payoff of $n-1-\epsilon$, for each player and any literal $l\in L$, either $l$ or $-l$ is played with  probability $\geq$ $\frac{1}{n} -2\epsilon-\frac{1}{ n^{2}}$ while the other is played with probability less than $\frac{1}{ n^{2}}$
\end{lemma}
\begin{proof}
 
	 If the row player plays $l$ and the column player plays $-l$, then according to the construction the sum of payoffs is $2n-8$.  It is not hard to see that the probability that this happens is less than  $\frac{\epsilon}{3} $ (by an argument similar to Lemma \ref{maxsw}). If both players do so with  probability greater than $\frac{\epsilon}{3} $, then the social welfare will be $2n-2-\frac{\epsilon}{3}(2n-2)-\frac{\epsilon}{3}(2n-8)=2n-2-2\epsilon$ and this will cause at least one player not gain a payoff of $n-1-\epsilon$. 
	
	Consider the literals $l$ and $-l$ from $L$ and assume without loss of generality, the row player plays $l$ more than $-l$.  Recall that we showed that each player plays either $l$ or $-l$ with probability of at least $\frac{1}{ n}-\epsilon-\frac{2 \epsilon}{n} \geq$ $ \frac{1}{ n} -2\epsilon$. For the sake of contradiction, let us assume that the row player plays $l$ with  probability less than $\frac{1}{n}-\left(\frac{1}{n^2}+2 \epsilon\right)$ (still greater than $\frac{1}{n^{2}}$)  which in turn, the row player has to play $-l$ with probability more than $\frac{1}{n^2}$.  The column player either plays $l$ with probability less than $\frac{1}{n^{2}}$ or plays $-l$ with probability greater than $\frac{1}{ n}-\epsilon-\frac{2 \epsilon}{n}$ .In either case, the probability that they  choose both $l$ and $-l$ is at least:
	
	\[\left[\frac{1 }{ n}-\left( \frac{1 }{ n^{2}} +2 \epsilon\right)\right]\left[\frac{1 }{ n^{2}}\right]= \frac{1 }{ n^{3}}-\frac{1 }{ n^{4}}-\frac{1 }{ 2n^{6}} \geq \frac{1}{2 n^{3}}=\epsilon\]

	This is impossible because we showed the probability that both outcomes happen, must be less than $\frac{\epsilon}{3}$.    So, if the row player must play a literal with  probability greater than $\frac{1}{n}-\left(\frac{1}{n^2}+2 \epsilon\right)$, the negation of this literal can be played with  probability of at most $\frac{1}{n^2}$. By a symmetric argument, the column player has to follow the same rules.
	
\end{proof}

\begin{remark}
    The analysis of the case that the column player plays $l$ with probability greater than $\frac{1}{n^2}$ is similar and was not provided in \cite{vadhan}. If the column player plays $l$ with  probability greater than $\frac{1}{n}-\frac{1}{n^2}-\epsilon$, by the fact that we know the row player plays $-l$ with a probability greater than $\frac{1}{n^2}$, the probability that they both play $l$ and $-l$ is at least $\epsilon>\frac{\epsilon}{3}$. If the column player plays $l$ with probability less than $\frac{1}{n}-\frac{1}{n^2}-\epsilon$, the column player has to play $-l$ with probability greater than $\frac{1}{n^2}$. By the assumption that we had, the row player plays $l$ with probability greater than $\frac{1}{2}(\frac{1}{n}-2\epsilon)$. The probability that both players play $l$ and $-l$ is still greater than $\frac{\epsilon}{3}$.
\end{remark}

\subsubsection{Proof of Proposition \ref{SVgame}} \label{SVproof}
The next step is proving Proposition \ref{SVgame}.

\begin{proof}
   If $\phi$ is satisfiable, we show that there is a uniform EXACT Nash equilibrium of $\mathcal{SV}(\phi,\epsilon)$ where each player plays $l^i \in L$ or $-l^i\in L$ (for all $i$) uniformly with  probability $\frac{1}{n}$.

		If $\phi$ is satisfiable, there exists an assignment where $v_1,\dots,v_n$ are assigned with \emph{true} or \emph{false} that satisfy all clauses. If $v_i$ is assigned \emph{true} then $l^i$, otherwise $-l^i$ will be selected as the strategies in the game. Then, $l^{1}, \dots, l^{n}$ are literals that correspond to a satisfying assignment. The expected payoff to each player is $n-1$ because they will always be playing one of $l$ or $-l$. Secondly, there are only two rules that give more than $n-1$, namely where one of the players plays either a variable or a clause.  For example, if the row player plays any clause $c$ with a positive probability, we know that the column player plays some literal in that clause $c$ with probability $\frac{1}{n}$ because the other player randomizes between literals in a satisfying assignment.  So in this case, the row player's payoff is at most $\frac{1}{n}\cdot 0+ \frac{(n-1)}{n}   \cdot n=n-1$, so, the row player gets the same payoff $n-1$ and is indifferent. This holds for variables as well. In conclusion, assuming one player plays uniform strategies on literals that satisfy the formula, the other player takes the same approach and this forms a Nash equilibrium with a guaranteed payoff of $n-1$.

Now suppose that $\phi$ is not satisfiable. We show that in any $\epsilon$-Nash equilibrium, at least one player always receives an expected payoff of less than $n-1-\epsilon$. For the approximate version, we need to do the following modification on the correspondence between literals and truth assignments compared to \cite{Conitzer}. We consider a literal is \emph{true} if it is played more often than its negation. If an assignment does not satisfy the formula, there is at least one clause that does not have a satisfying literal. Any of the players will turn to this clause strategy to receive a payoff of $n$ whenever the opponent plays a literal that is not in that clause. We know that the column player plays literals with probability $>1-\epsilon$ by Lemma \ref{maxsw}, and there are only $3$ literals in each clause each of which the column player plays with probability $\leq \frac{1 }{n^{2}} $ by Lemma \ref{bothliterals}. By changing the strategy to this clause, the row player will receive at least $\left(1-\epsilon-\frac{3}{n^{2}} \right) n>n-1+2 \epsilon$. So either the row player can do $\epsilon$ better by changing his strategy or he is already receiving $n-1+2\epsilon$ and so the other player does not have a guaranteed payoff of at least $n-1-\epsilon$.   
\end{proof}

\subsection{Hardness of Partition Nash and Proof of Proposition \ref{expartiiton}}

\begin{proof}
As stated, it is easy to check whether the two conditions that we need for a partition Nash are satisfied or not. So, this problem is in NP.
 Proposition \ref{modifedconitzer} shows that even the POLY-approx version of is NP-hard. We know that in $\mathcal{G}(\phi,\epsilon)$, by Proposition \ref{modifedconitzer}, any $\epsilon$-Nash equilibrium that does not have $f$ in the support of both players, has the following property. For each player $i\in \{1,2\}$ and any specific literal $l_i$ such that $\mathrm{g}(l_2)=\mathrm{g}(l_1)=l$, one of $l_i\in L_i$ or $-l_i\in L_i$  is played with probability greater than $\frac{1}{n}-\frac{1}{n^2}-2\epsilon$ while the other is played with probability smaller than $\frac{1}{n^2}$. All other possible strategies are played with a probability of at most $\epsilon$ since literals have to be played with probability $1-\epsilon$.

In $\mathcal{G}(\phi,\epsilon)$, any EXACT disjoint Nash equilibrium should have the property that for all $l_1\in L_1$ and $l_2\in L_2$ such that $g(l_1)=g(l_2)=l$, one of $l_1$ or $-l_1$ for the row player and one of $l_2$ or $-l_2$ should be played with probability $\frac{1}{n}$ while the other has to be played with probability zero directly followed by \cite{Conitzer}. Note that if one of the players plays $f$, the other player has to play $f$ too which we are not concerned with in this proof. In this case, all players will play exclusively from their associated literals since all other literals are dominated by all other strategies.

If the formula $\phi$ is satisfiable, an EXACT disjoint Nash can be generated by \ref{modifedconitzer}. Assume that $(x^*,y^*)$ is an EXACT disjoint Nash equilibrium of $\mathcal{G}(\phi,\epsilon)$. We modify this strategy to $(x^-,y^-)$ and generate a $\epsilon$-partition Nash equilibrium. Consider $\epsilon^\prime=\frac{\epsilon}{cn}$ and recall that $\epsilon=\frac{1}{2n^3}$. The reason that we chose $\epsilon^\prime$ to be very small is that we have some strictly dominated strategies ($c$ is large enough to cover all these rules for a partition) and also $f$ cannot have a probability greater than $\frac{\epsilon}{n}$. We take a small fraction $\frac{\epsilon^\prime}{2}$ of one pure strategy from the column player and distribute it equally to strategies that are not in the support of the row and the column player for the column player.  It is easy to see that $(x^-,y^-)$ is an $\epsilon$-approximate partition Nash equilibrium by Lemma \ref{blowup}. Now, both properties of a partition Nash instance are satisfied.  It is obvious that if we have an approximate partition Nash $(x^-,y^-)$, we simply can generate an EXACT disjoint Nash equilibrium that can imply a satisfying assignment. This is because all literals $l$ that are played with probability greater than $\frac{1}{n}-2\epsilon-\frac{1}{n^2}$ can have probability $\frac{1}{n}$ can determine an assignment of the formula $\phi$.

\end{proof}
\begin{remark}
Proof of \ref{itemwise} is simply followed by dividing the strategies that are not related to literals that provide an answer to SAT equally to both players.
\end{remark}

\section{Diagonally Modified Games and Constrained Far Equilibrium}

\subsection{Proof of Proposition \ref{zerodiag}}

We know that essentially, there exists a game $\mathcal{G}$ that is known to be hard for finding an equilibrium in \cite{Chen} even for inverse polynomial approximation as we restate it in the following:

		\begin{theorem}[\cite{Chen}]
		    For any $c>0$, the problem of computing an $n^{-c}$-approximate Nash equilibrium of a two-player game is PPAD-complete. 
		\end{theorem}

This game can be reduced to a diagonally modified game by using the construction that we provide in the following. Suppose this normal form game $\mathcal{G}$ has payoff matrices $U_1$ and $U_2$ where both players' strategies come from $S=[n]$. In the following game $\mathcal{D^\text{M}(G)^{\prime}}$ that we generate (which is also a diagonally modified game), any Nash equilibrium of the original game $\mathcal{G}$, forms a disjoint Nash equilibrium (which is also a constrained disjoint equilibrium) in the generated game. Note that for any Nash equilibrium in this game, the row player can only assign positive probabilities on $S_1$ and this holds for the column player with $S_2$ respectively. The proof follows by Example \ref{changelabellemma}.
\[   
U^{\prime}_1(x,y)= 
\begin{cases}
	U^{\mathcal{G}}_1(i,j) & \quad i\in S_1 \thickspace and \thickspace j\in S_2 \\ 
    -M  & \quad otherwise \\
	
\end{cases}
\]
\[ 	
U^{\prime}_2(i,j)= 
\begin{cases}
	U^{\mathcal{G}}_2(i,j) & \quad i\in S_1 \thickspace and \thickspace  j\in S_2 \\ 
	-M & \quad otherwise \\
	
\end{cases}
\]		

 Both diagonal members of $U^{\prime}_1$ and $U^{\prime}_2$ can be arbitrary negative numbers that must be strictly smaller than the minimum entry (generally zero) in the original game. Any (EXACT)  Nash equilibrium that is found in the first game can be mapped to a disjoint Nash equilibrium of the second game by using the associated copy of each strategy for each player. It is obvious that any constrained disjoint equilibrium (that is also a disjoint Nash equilibrium)that is found in the second game can be mapped to a Nash equilibrium in the first game as well.

    Now suppose that we are given an approximate constrained disjoint Nash equilibrium $(x^*,y^*)$  in $\mathcal{D^{\text{M}}(G)}$ with approximation error $\epsilon$. In a  $\epsilon$-constrained disjoint equilibrium of $\mathcal{D^\text{M}(G)^{\prime}}$ the players can play strategies that are not associated with them with probability smaller than $\frac{\epsilon}{M}$. By applying Lemma \ref{blowup}, we can generate a valid approximate Nash equilibrium for $\mathcal{G}$.

\subsection{Proof of Proposition \ref{noescape} }

      If we have a $\theta$-restricted  disjoint equilibrium $(x^*,y^*)$ for $\mathcal{G}$, we prove that this equilibrium will be a disjoint Nash equilibrium for $\mathcal{D^{\text{M}}(G)}$. For all pure strategies $s\in S$, we need to prove:
\[U^{\mathcal{D^{\text{M}}(G)}}_1(x^*,y^*) \geq U^{\mathcal{D^{\text{M}}(G)}}_1(s,y^*)\]

We can easily show that this inequality is correct for all $s\notin Supp(y^*)$ since we know that $U^{\mathcal{D^{\text{M}}(G)}}_1(x^*,y^*)=U^{\mathcal{G}}_1(x^*,y^*)$ and $U^{\mathcal{G}}_1(s,y^*)=U^{\mathcal{D^{\text{M}}(G)}}_1(s,y^*)$. 
Next, consider any $s\in Supp(y^*)$.  In $(x^*,y^*)$, it is impossible to have probabilities smaller than $\frac{1}{M}$ except zero. Then, can conclude that $U^{\mathcal{D^{\text{M}}(G)}}_1(s,y^*)< 0$. This follows by the fact that $s$ is played with  probability $1$ and the column player has to $s$ with probability greater than $\frac{1}{M}$.

We know that the payoff of $(x^*,y^*)$ is positive in $\mathcal{D^{\text{M}}(G)}$ which means it is impossible to deviate to $s$ in this case. This is correct for the column player by a symmetric argument. This also holds for the approximation notion by a similar argument.

\subsubsection{Constrained Disjoint Equilibrium is in NP}\label{innpfar}

\begin{proposition}\label{tresp}
The problem of finding a constrained disjoint equilibrium in a bi-matrix is in $NP$.
\end{proposition}

\begin{proof}
We use the equivalent matrix format to calculate the expected payoffs. Suppose that we have a constrained disjoint equilibrium $\left(x^{*}, y^{*}\right) \in \triangle_{n} \times \triangle_{n}$. We can check the following for the row (similarly for the column player):
	\[	 \forall i \in\{1, \dots, m\} \cap  S-Supp(y^{*}), x^{* T} R y^{*}  \geq e_{i}^{T} R y^{*}
\]

	Suppose $x=(\alpha_{i_1},\dots,\alpha_{i_m})$ is an arbitrarily mixed strategy in $\mathcal{R}^n_{disjoint}(y^*)$ where we removed all strategies that are assigned with a zero probability. We show that any strategy in $\mathcal{R}^n_{disjoint}(x^*)$ could be generated as follows. We multiply $\alpha_i$ for each $i$ to the previous inequality and calculate the summation as follows:
	\[
	\begin{aligned}
		\sum_{i=1}^{m} \alpha_i x^{* T} R y^{*} & \geq \sum_{i=1}^{m} \alpha_i e_{i}^{T} R y^{*}
	\end{aligned}
	\]
	
	Since $x$ is a distribution, we have $\sum_{i=1}^{m} \alpha_i=1 $. And the right hand side will be  $\sum_{i=1}^{m} \alpha_i e_{i}^{T} R y^{*}=x R y^{*}$. Finally, the result will be:

	\[
	 \forall x \in \mathcal{R}^n_{disjoint}(y^*),~ x^{* T} R y^{*} \geq x^{T} R y^{*}
		\]
	
	And for the second player we can take the same approach:
		
	\[
\forall y \in \mathcal{R}^n_{disjoint}(x^*),~	x^{* T} C y^{*} \geq x^{* T} C y
		\]
	
\end{proof}

\begin{corollary}
   The problem of finding an approximate restricted  disjoint equilibrium in a bi-matrix game is in $NP$.

\end{corollary}

\begin{proof}
    The restricted version of constrained disjoint has one additional condition where checking the condition can be done in polynomial time. 
\end{proof}

\end{document}